\input{psfig.tex}
\documentstyle[10pt,emulateapj]{article}

\def\solm{M$_{\odot}\,$} 
\def\solmm{{\rm M}_{\odot}\,} 
\def\kms{km s$^{-1}$}
\def\ls{\mathrel{\hbox{\rlap{\hbox{\lower4pt\hbox{$\sim$}}}\hbox{$<$}}}}
\def\gs{\mathrel{\hbox{\rlap{\hbox{\lower4pt\hbox{$\sim$}}}\hbox{$>$}}}}

\begin{document}
\submitted{Accepted to the Astronomical Journal}
\title{A Direct Measurement of Major Galaxy Mergers at $z \ls 3$}

\author{Christopher J. Conselice$^{1,2}$, Matthew A. Bershady$^{3}$, Mark 
Dickinson$^{4}$, Casey Papovich$^{5}$}

\altaffiltext{1}{California Institute of Technology, Mail Code 105-24, Pasadena CA, 91125}
\altaffiltext{2}{NSF Astronomy \& Astrophysics Postdoctoral Fellow}
\altaffiltext{3}{Department of Astronomy, University of Wisconsin, Madison 
475 N. Charter St. Madison, WI, 53706-1582}

\altaffiltext{4}{Space Telescope Science Institute, 3700 San Martin Drive, 
Baltimore, MD. 21218}

\altaffiltext{5}{Department of Astronomy, University of Arizona}

\begin{abstract}

This paper presents direct evidence for hierarchical galaxy assembly out to 
redshifts $z \sim 3$. We identify major mergers using the model-independent 
CAS (concentration, asymmetry, clumpiness) physical morphological system on 
galaxies detected, and photometrically selected, in the WFPC2 and NICMOS 
Hubble Deep Field North.  We specifically use the asymmetric distributions of 
rest-frame optical light measured through the asymmetry parameter ($A$) to 
determine the fraction of galaxies undergoing major mergers as a function of 
redshift ($z$), stellar mass (M$_{\star}$), and absolute magnitude 
(M$_{\rm B}$).  We find that the fraction of galaxies consistent with 
undergoing a major merger increases with redshift for all galaxies, but most 
significantly, at 5 - 10 $\sigma$ confidence, for the most luminous and 
massive systems.  The highest merger fractions we find are 40\% - 50\% for 
galaxies with M$_{\rm B} < -21$, or M$_{\star} > 10^{10}$ \solm at $z > 2.5$, 
i.e., objects identified as Lyman-break galaxies.  Using these results, we 
model the merger fraction evolution in the form: 
f$_{m}$($A$,M$_{\star}$,M$_{\rm B}$, $z$)~=~f$_{0}~\times (1+z)^{m_{A}}$. We
find $m_{A}$ values $\sim 4-6$ for the most luminous and massive galaxies, 
while lower mass and less luminous galaxies have smaller $m_{A}$ values. We 
use these merger fractions, combined with merger time scales calculated from 
N-body simulations, to derive galaxy merger rates to $z \sim 3$.  We also use 
stellar masses of HDF-N galaxies to determine the mass accretion rate of 
field galaxies involved in major mergers.  We find an average stellar mass 
accretion rate of $\dot{\rm M}_{\rm G} \sim 4 \times 10^{8}$ 
\solm Gyr$^{-1}$ galaxy$^{-1}$ at $z \sim 1$ for galaxies with stellar masses 
M$_{\star} > 10^{9}$ \solm.  This accretion rate changes with redshift 
as: $\dot{\rm M}_{\rm G} = 1.6 \times 10^{8}$ $(1+z)^{0.99\pm0.32}$ 
\solm Gyr$^{-1}$ galaxy$^{-1}$.   We also find that the fraction of stellar 
mass density in galaxies involved in major mergers increases with redshift, 
with a peak mass fraction $\sim 0.5$ for the brightest, M$_{\rm B} < -21$, 
and most massive, M$_{\star} > 10^{10}$ \solm, systems near $z \sim 2.5$. By 
comparing merger fractions predicted in Cold Dark Matter semi-analytic 
models with our results we find a reasonably good agreement for the largest 
and brightest systems, although we find more low-mass galaxy mergers at 
lower redshifts than what these models predict.

\end{abstract}

\section{Introduction}

There are two principle processes in galaxy formation: the
assembly of mass, both baryonic and dark, through accretion and mergers, and 
the conversion of baryons into stars.  While the latter process is now 
statistically mapped out to nearly the beginning of the universe using a 
variety of techniques (e.g., Lilly et al. 1996; Madau et al. 1998), 
we are just beginning to understand the process and history of mass assembly.
Furthermore, there are several types of mass assembly that are related in an 
unknown way, including the collapse, infall, and accretion of dark and baryonic
matter, which are likely
related to how baryonic material is converted into stars.    Assembly
of galaxies and dark halos through mergers and accretion is also
potentially a major player in black hole and AGN evolution, the production
of gravitational waves, the triggering of star formation, and possibly a 
driver of supernova and gamma-ray burst rates.

Hierarchical mass assembly is also the cornerstone of all Cold
Dark Matter (CDM) models of galaxy formation (e.g., Cole et al. 2000
and references therein). These currently favored models clearly predict
that dark halos of modern galaxies were formed in the past through the
process of repeated merging of, and buildup from, smaller systems.
Although CDM models predict that dark halos merge, it
is not clear if galaxy formation, or star formation, occurs during, before,
or after dark-halo mergers. If baryons collapse to form stars in dark halos 
before
a significant amount of halo merging, then, based on dynamical friction
arguments, we should witness mergers of galaxy stellar components. On the 
other hand, it is also 
possible that gas cools and forms stars after dark halos merge, producing
the large galaxies we see today (e.g., Noguchi 2000).
 
One method of determining if and how galaxies form by merging 
is to directly measure the fraction of galaxies undergoing mergers, and mass 
assembly occurring by mergers, at various look-back 
times and estimate from these merger and mass assembly rates.  While the 
star formation history of the universe is retrievable in part
by examining nearby resolved galaxy stellar `fossil' populations, the mass 
assembly 
history for nearby galaxies is mostly lost through equilibrium and relaxation 
processes.  Some merger tidal debris may remain for several Gyrs, such
as in our own galaxy (e.g., Ibata et al. 2002; Newberg et al. 2002), although
these are likely the result of recent minor mergers.
There is also considerable observational 
evidence for recent major mergers in the
local universe (e.g. Schweizer \& Seitzer 1988; Borne et al. 2000) and 
accretion of low-mass galaxies onto larger ones (Zaritsky \& Rix 1997).  
However, direct evidence for an increase in the galaxy merger rate
at high redshift has not yet been established, despite the considerable
circumstantial evidence. This includes
evolving luminosity functions (e.g., Lilly et al. 1995; Ellis et
al. 1996), and the appearance of distant `irregular' galaxies seen in HST WFPC2
images (e.g., Driver
et al. 1995; Glazebrook et al. 1995; Abraham et al. 1996). Indeed an
enhanced past merger rate can plausibly explain both the "faint-blue-galaxy"
excess, and may well be the physical mechanism driving the
evolution of the blue galaxy luminosity function seen at intermediate
redshifts. This idea was suggested over a decade ago (Broadhurst,
Ellis \& Glazebrook 1992), but direct evidence for evolution in the galaxy
merger rate has not yet been established.

The most popular method for measuring
the evolution of galaxy mergers at high-redshift is through pair counts 
(e.g., Zepf \& Koo 1989; Burkey et al. 1994; Patton et al. 1997;  Wu \& Keel 
1998; Le~F\'{e}vre et al. 2000) or kinematic
pairs (Carlberg et al. 2000).  After correcting for selection effects and 
biases (e.g. Patton et al. 2002), pair-count methods can be used to study
major galaxy mergers out to $z \sim 1$. At high redshifts, however, this 
methods becomes difficult
and expensive in telescope time, due to the many redshifts needed.  There has 
in fact
never been a measurement of merger fractions, or merger rates, at redshifts 
$z > 1$, although detections of high redshift galaxy merging have been 
claimed (e.g. Neuschaefer et al. 1997).

Perhaps the best way to understand and characterize the merging process
is to observe high redshift galaxies and determine which are undergoing
mergers based on their stellar light distributions.   We argue in
this paper, and in Conselice (2003), that this can be done using 
the observed structures of galaxies. At high redshifts, we can do this
using the {\em Hubble Deep Fields} (HDF) 
(Ferguson, Dickinson \& Williams 2000).  Conselice et al. (2000a) (hereafter
CBJ00), Conselice et 
al. 2000b and Conselice (2003) argue that galaxies 
undergoing major mergers can be identified through their large structural 
asymmetries, one aspect of the CAS (Concentration, Asymmetry, Clumpiness)
morphological system (Conselice 2003).

Previously, a color-asymmetry diagram was used by Conselice \& Bershady (1999)
to determine the fraction of galaxies undergoing mergers 
in the HDF, based on the original WFPC2 images, finding a merger
fraction of 40\%.   These previous studies are however potentially
biased by morphological K-corrections, where a galaxy's appearance in
the rest-frame UV is not necessarily similar to its 
rest-frame optical morphology.  This problem is removed in this paper
through the use of HDF NICMOS images, where the rest-frame optical light of 
galaxies is sampled out to $z \sim 2.5$.

By using the CAS system on galaxies found in the Hubble Deep Field North 
(Williams et al. 1996), and after understanding systematics 
and biases through simulations, we are able to measure the major merger history
of galaxies out to $z \sim 3$.   We use rest-frame B-band asymmetries with 
other structural and photometric indices such as radii, absolute magnitude 
(M$_{\rm B}$), and stellar masses (M$_{\star}$) to further address the 
question of how
galaxy mergers have evolved over time.   Using this method we find that out to 
$z \sim 1$ the fraction of galaxies involved in major mergers
increases, as has previously been noted (e.g., LeFevre et al. 2000;
Patton et al. 2002).   We argue that the fraction of galaxies undergoing 
mergers is lower for the fainter
and lower mass galaxies at higher redshifts.  The merger fraction
does however continue to increase with redshift for the brightest and most 
massive systems.  Based on this, we compute and quantitatively characterize
the merger and mass assembly
rates of galaxies due to merging, out to $z \sim 3$.   We also investigate
how the fraction of the total stellar mass density in galaxies involved in 
major mergers changes with time, magnitude, and mass, and compare our 
results to predicted values from Cold Dark Matter models. 

This paper is organized as follows:  \S 2 is an explanation of the 
data, including redshift information and how the asymmetry parameter is 
measured for HDF galaxies. In \S 3 we briefly explain the basis
for our argument that structures of galaxies holds information from which
a merger origin can be derived, \S 4 gives the basic results of the asymmetry
measurements and how they correlate with other physical properties, including
absolute magnitude, color, and stellar mass, to argue 
that we can determine the merger history of galaxies out to $z \sim 3$. \S
4 further
explores the comparison of merger histories with Cold Dark Matter (CDM)
models and \S 5 is a summary.  The cosmology H$_{0} = 70$ km s$^{-1}$ 
Mpc$^{-1}$, $\Omega_{\rm m} = 0.3$, and $\Omega_{\lambda} = 0.7$ is used 
throughout this paper.

\section{Photometric Data and Parameters}

\subsection{Imaging Data}

The images
we use for our morphological analyses were acquired with the Hubble Space 
Telescope as part of the Hubble Deep Field North optical and near infrared
campaigns. Optical data from the Wide-Field Planetary Camera-2 (WFPC2) of the
HDF-North (Williams et al. 1996) are combined with Near-Infrared (NIR)
observations of the same field taken with the Near Infrared Imaging Camera and
Multi-Object Spectrometer (NICMOS) (Dickinson 1998, Dickinson et al. 2000). 
The optical data consist of 
images in the WFPC2 filters: F300W (U), F450W (B), F606W 
(V) and F814W (I).  The NICMOS HDF-North images were taken in June 1998 with
observations in the F110W (J) and F160W (H) near 
infrared bands. We also use a deep K-band image obtained with the Kitt
Peak Mayall 4m.

From these images a total of 1212 galaxies in the HDF were detected with 
SExtractor at AB magnitudes brighter than J = 27,  based on the photometry of
Dickinson et al. (2000).  We compute magnitudes, 
and radii (\S 2.3) for these galaxies within segmentation maps produced 
by SExtractor (Bertin \& Arnouts 1996).  These galaxies
constitute the sample used throughout this paper, although we effectively
use only a fraction of these as we only consider galaxies with 
M$_{\rm B} < -18$ to avoid strong selection effects and biases (\S 2.3).

The shapes, sizes,
structures, and morphologies of HDF galaxies, as observed in these six
pass-bands, are affected by instrumental effects, such as point 
spread functions, in addition to the intrinsic
morphological K-corrections and surface brightness dimming.
The angular resolution of the NICMOS HDF images is poorer than that of
the WFPC2 data because of the longer wavelength diffraction limit and
the pixel undersampling of NICMOS Camera 3.  The effective PSF FWHM
is approximately 0.22''.  To minimize systematic errors
associated with comparing galaxies at different resolutions
(CBJ00), we have convolved the WFPC2 images to match the NICMOS PSF,
and made all morphological measurements from these PSF--matched data.
Note also that the angular diameter distance to galaxies varies by less
than 40\% over the redshift range $0.5 < z < 3$, and by $< 20$ \% for
$0.7 < z < 3$, where the vast majority of the galaxies analyzed in this
paper lie.  Therefore the resolution of our data in linear physical
units (e.g., kpc) is nearly constant over the redshift and wavelength
ranges considered here.

\subsection{Redshifts, Photometry and Stellar Masses}

Photometry is done in all six HDF passbands, and the K-band image, for every 
galaxy, with photometric redshifts derived based on observed spectral 
energy distributions (Budavari et al. 2000).  We
use these photometric redshifts and spectroscopic ones, when available, to
compute the rest-frame absolute B-band magnitude, M$_{\rm B}$, and rest-frame 
Johnson (B$-$V) color for each galaxy.  Rest frame magnitudes and colors are 
computed by
fitting the spectral energy distributions of each galaxy to templates.
The best fit template is then used to calculate the (B$-$V) color of each 
galaxy based on interpolation of its observed SED.  A total of 157 
spectroscopic galaxy redshifts were 
available to us from Cohen et al. (2000),  Dawson et al. (2001) 
and K. Adelberger/C. Steidel (priv. comm).  Figure~1 shows the absolute
magnitude distribution of all galaxies to our magnitude limit of J = 27, 
with derived (B$-$V) colors, plotted as a function
of redshift ($z$).  

The stellar masses used in this paper come from the
analysis of Papovich et al. (2001), Papovich (2002) and Dickinson et al. 
(2003).
The stellar masses are derived using the seven band photometry described
above and by assuming an initial mass function (IMF), metallicity, and a
star formation history. The
star formation history can be modeled, for example, as instantaneous, 
``bursty'', or exponentially
declining.  While there are several different possible initial mass
functions (IMF) and star-formation histories for these galaxies,
we use the model results of Papovich (2002) where the IMF is 
Salpeter, the metallicity solar, and the star
formation history of each galaxy is monotonic.  The stellar masses
are also not computed for all the galaxies in the HDF, as many 
are too faint for a reliable computation. 

For this particular choice of IMF, metallicity and star formation history,
the typical uncertainties on the stellar masses are about a factor of two
(Papovich et al. 2001, Papovich 2002).  The uncertainties are larger at
higher redshifts, where the photometric data does not reach red rest--frame
wavelengths, and for the bluest galaxies, where varying amounts of stellar
mass can be hidden by the bright light of ongoing star formation.

\subsection{Asymmetries}

\subsubsection{Computing Asymmetries}

The main quantitative structural parameter used in this paper is the 
asymmetry ($A$) index, one property of the CAS physical morphology system 
(Conselice 2003).
Asymmetry has been used previously as a morphological parameter for nearby 
(Conselice 1997; CBJ00) and distant galaxies (e.g. Shade et al. 1995; 
Abraham et al. 1996; Conselice \& Bershady 1999). The asymmetry 
computation we use is described in detail in CBJ00.  The
basic computation of $A$ involves rotating and subtracting a galaxy image
from itself, and comparing the summation of the absolute value of these 
residuals to the original galaxy's flux.

Our computation differs from the prescription set out in CBJ00 through our 
use of the SExtrator dimensionless, `Kron radius multiplier' (k$_{r}$), times 
a factor of the semi-major axis ($a$), to define the area within which 
asymmetries are computed.  The
SExtractor Kron radius multiplier (k$_{r}$) is a dimensionless quantity
used to scale the ellipse whose semi-major (a) and semi-minor (b)
axes are defined by first moments of the light profile.
This is different from the prescription of CBJ00 where the Petrosian (1976)
$\eta$ radius was used.   However,
comparisons of the $\eta$ radius used in CBJ00
and the first moment radii for bright HDF galaxies reveals that the maximum 
difference of k$_{r} \times a$ (hereafter Kron radius) and the Petrosian 
radius is very small, usually 
a few pixels ($\sim 0.5$\arcsec) at most, and always different by less than 
10\%.  Asymmetries 
computed using the two methods are also indistinguishable for 
galaxies in the HDF.  Simulations
also show that defining the radius in this manner is robust when a galaxy
is moved to higher redshifts (\S 2.3.5).

Asymmetries are computed for all galaxies 
in each observed WFPC2 and NICMOS band.
The rest-frame B-band asymmetries are then derived using the spectroscopic or 
photometric redshift of each galaxy, and then interpolating rest-frame
B-band asymmetries using
the observed asymmetries.  The rest-frame B-band 
asymmetry is computed in this way 
through a linear combination of the asymmetry values in the two filters 
nearest the rest-frame B-band wavelength.  For galaxies at $z > 2.6$ where the 
central rest-frame B-band filter wavelength becomes higher than the central 
wavelength of the H-band, we assume A(B)$_{(\rm rest)}$ = A(H)$_{(\rm Obs)}$. 

There are, however, many biases, random, and systematic errors that must be 
understood in detail, and accounted for, before we
can reliably interpret and use these asymmetry values to derive evolution.  
The general procedure for measuring asymmetries, as outlined above, relies on 
a robustly defined center, radius, and background area.  As the Hubble Deep 
Field North is crowded with galaxies, but not confusion limited, finding 
suitable background areas to measure sky statistics is not trivial.  We 
examine this problem, and the centering and radius issues later in this paper, 
although see CBJ00 
for a detailed description of these problems using nearby galaxies.
To address some of these issues, we examine a subsample of the 38 
brightest HDF galaxies, with
M$_{\rm B} < -18$, at redshifts 0.4$<z<$0.7
in the HDF to determine how centering routines, differing radii and 
background subtraction methods affect the measured asymmetries.  We also
measure asymmetries within radii defined within the SExtractor region, and
we  investigate possible galaxy contamination in \S 3.3.

\subsubsection{Asymmetry Systematics with Radius}

The problem with
choosing a radius to measure asymmetries, and other structural parameters, 
lies in the trade-off between 
galaxy coverage and noise, including contamination from other
galaxies.  CBJ00 determined that the larger the radius, the more representative
the computed
asymmetry index is in comparison with a fiducial 'total' asymmetry, and the 
better its asymmetry values correlate with other physical properties.   On the
other hand, if 
radii used for computing asymmetries are large, then more background
will be present in the computation, which increases the noise on the asymmetry
measurement.  A larger radius also increases the chance of contamination from
other galaxies, and thus requires careful attention to neighboring objects.

While there is no fool-proof way of determining the best radius to
measure asymmetries, we can constrain this to some extent by
examining asymmetries of the relatively nearby galaxy sample in the HDF
computed at various factors of the Kron radii.   These measurements are
plotted in 
Figure~2 as a function of the Kron radii as measured by SExtractor.    The 
asymmetry parameter is
measured for the 38 nearby galaxies in the HDF at five different radii: 
0.25~k$_{r} \times a$,  0.5~k$_{r} \times a$,  1.0~k$_{r} \times a$, 
1.5~k$_{r} \times a$, and 2.0~k$_{r} \times a$.  The average and $\pm$1$\sigma$
variations of the asymmetries for these 38 galaxies at these 
radii are: 0.08$\pm$0.06, 0.13$\pm$0.11,
0.24$\pm$0.26, 0.34$\pm$0.31, and 0.44$\pm$0.33. 

As for nearby galaxies
(e.g., CBJ00), asymmetries increase at higher radii. Some of the higher
asymmetries at larger radii are the result of contamination from nearby
galaxies. We also find that errors on individual asymmetry 
measurements increase when using larger measuring radii.  
To strike a balance between representative radius and noise on
individual measurements, we use
the radii 0.5 $\times$ k$_{r} \times a$ to measure asymmetries. This is
also the radius that matches the apparent radius when estimated by
eye.  Through
tests we found this radius to be similar to the r($\eta = 0.2$) radius
used to measure asymmetries for nearby galaxies, and these radii 
can be robustly measured
for galaxies out to redshifts of $z \sim 3$ (\S 2.3.5).

\subsubsection{Search Radii}
 
CBJ00 found that asymmetries do not significantly depend
upon the search size used to find the minimum asymmetry. The
search size is the pixel size that the asymmetry computation code
uses to find the minimum asymmetry by computing $A$ values at centers that 
differ by the search size (CBJ00) until the minimum asymmetry is found.   We 
tested this on the HDF data using several different search radii, and always
found the same asymmetry, independent of search size.  For example, when
changing the search radius from 0.5 to 0.1 pixels, the values of the 
asymmetries only change by $\delta A = 0.04$ at most.

\subsubsection{Background Measurement Systematics and Effects of
Correlated Noise}

The method of background removal is a very important issue that we
address here in some detail.  The asymmetry computation as outlined in 
\S 2.3.1
needs to be corrected for background noise, and this is usually done
using a patch of sky in the image that contains no part of any galaxy.
This approach is necessary, as an annulus surrounding a galaxy may be 
contaminated by faint outer regions which have dimmed below the noise level 
(e.g., the Tolman (1934) effect).

The background noise level in our images varies
from place to place in the NICMOS images, and to a lesser extent the
WFPC2 data, due to variations in quantum efficiency and dark current
noise over the detector array.   For this reason, no one ``blank spot''
used to constrain background noise in the asymmetry measurements can
be considered wholly representative.  To determine the degree to which
this may affect the asymmetries, we show in Figure~3 the asymmetries of all
the 1212 galaxies measured using five
different background regions throughout the HDF images (listed
as asymmetry runs 1 through 5).  These five
background regions are at the same physical place in each band, and were 
carefully chosen to span the area of the HDF.  We
only show the resulting asymmetry distributions in Figure~3 for the I and H 
band asymmetries, but these are representative of the B,V and J band 
asymmetries, respectively.  

As can be seen, the I-band asymmetries are relatively constant using
the different backgrounds, but the H-band asymmetries can and do vary.  
To overcome this problem, the effective asymmetry measured
in each band is found by averaging the asymmetries for the five different
backgrounds.  The 1$\sigma$ variation in these measurements is added in
quadrature to the average asymmetry measurement error, and this value is used
as the asymmetry error.

We also performed a series of simulations to determine the effects of
correlated noise on the measurement of asymmetries.  This was done by
creating fake noise maps and placing galaxies into them, and then measuring
their asymmetries.   When we correlate the noise by
smoothing the initial noise map by some filter and then remeasure the 
resulting asymmetries, we find that the variation in $A$ is tiny, $\delta$A
$\sim \pm 0.03$.  This is due to the fact that we empirically remove the 
background by using a patch of sky, which in principle matches the noise
pattern under the galaxy being studied. 

\subsubsection{Detection and Simulations of Asymmetry Variations with Redshift}

A significant issue that must be addressed in any study that compares
properties of galaxies at different redshifts is the fact that measured
properties, and the detection of galaxies themselves, change solely due
to redshift effects.  The rapidly increasing luminosity distance of 
galaxies, with the slowly changing angular size distance,
produces a $(1+z)^{4}$ decline in surface brightness.  The result 
is that objects we
detect in the HDF at lower redshifts might not be detectable at higher
redshifts, thereby invalidating some comparisons.   Changes in S/N
and resolution due to redshift can also mask, or mimic, real evolution.
We addresses these issues using simulations 
and apply this information to correct our asymmetry measurements, and 
to understand our detection completeness.  Note that we always use
interpolated rest-frame B-band values of each galaxy, thus we remove all 
morphological k-corrections.

Simulations were carried out by using the 38 galaxies at 
0.4$< z <$0.7 which have M$_{\rm B} < -18$ as described in \S 2.3.1.
The images of these objects in their approximate rest-frame
$B$-band (observed F606W, or $V$-band), are simulated
as they would appear at various redshifts from $z = 1$ to 3 in
the redder HDF filters (at $z > 3$ we can no longer sample
rest-frame $B$ morphologies).  These simulations were done by
creating the background for each of the HDF bands, with the
same noise characteristics, then randomly placing the simulated galaxies
into these backgrounds.  The galaxies are reduced in resolution, signal to
noise, flux and surface brightness, and convolved with either the NICMOS
or WFPC2 PSF (see Conselice 2003). 

After these galaxies were simulated, we ran the SExtractor detection software 
using the exact same criteria
used for the original HDF detections (Dickinson et al. 2003b in prep). From 
this, we 
are able to determine the detection completeness at each simulated redshift, 
and measure
the asymmetries and radii of the galaxies detected.  Doing this allows us to
better understand the systematics produced by non-detections and how
asymmetry changes due to cosmological effects as opposed to real evolution.
We simulate this local sample of HDF galaxies to redshifts
$z = 1,~1.5,~2,~2.5,~3,~4,~5, {\rm and}~6$.  While we do not discuss real HDF
galaxy asymmetries for objects above $z \sim 3$ in this paper, we include 
these higher redshift simulations for completeness, and to investigate
trends with redshift.

The results for these simulations are shown in Figure~4 and are quantified in
Table~1.  First, all of the 38 objects we simulate from low$-z$ remain 
detectable
until $z \sim 2.5$ when the completeness begins to drop (Figure~4).    
The measured values of the asymmetries generally become
lower at higher redshifts (Figure~4).    The average corrections necessary to 
account for these effects are listed in Table~1 and are generally low, with 
differences 
$\delta A$ = 0.04 - 0.06 for the redshift ranges studied in this paper.

There is also only a very slight difference in retrieved asymmetries 
for faint galaxies with different magnitudes in each redshift range.
The fainter galaxies at each redshift range are generally affected by 
noisie more and hence the systematics effects (and corrections) are
larger, with values $\sim 0.02$ higher, on average, than the
brighter systems.  As this is usually smaller than the random measurement 
errors for these faint galaxies, we do not account for
this small difference.
As discussed in CBJ00, where similar simulations are done in terms of S/N
ratio and resolution, the asymmetry index is not greatly affected by the 
reduced resolution and lower
S/N ratios for galaxies with M$_{\rm B} < -18$.  
Figure~4 does show however that by using
a magnitude limit of M$_{\rm B} = -18$ we begin to become incomplete at
redshifts $z > 2.5$.  

Our radii, defined as 0.5$\times$ k$_{r} \times a$ (\S 2.3.1), also shows
a slight decrease when the 38 low redshift galaxies are simulated to higher
redshift and then redetected with SExtractor (Figure 4b).  For the
remainder of the paper we use these results to apply 
slights corrections (Table~1) to the measured asymmetries and 
radii for galaxies found in the HDF at $z > 0.7$.  

\subsubsection{Final Asymmetry Values and Errors}

The final asymmetry
for each galaxy in every band is computed by taking the average of 
the asymmetry values computed using the different backgrounds (\S 2.3.4), and 
depending on the redshift, applying a systematic error correction as described
in \S 2.3.5.  The
error of each asymmetry measurement is computed by combining the average
measured error with the RMS of the asymmetry values.  The average
errors on our asymmetry values remain extremely low for the brightest 
galaxies at M$_{\rm B} < -21$
with $<\delta A> = 0.04$ out to $z \sim 3$ (Table~2).   
All of the random errors in fact remain
rather low, except for the faintest galaxies, with M$_{\rm B} > -19$, at
$z > 2.5$ where the errors approach $<\delta A> \sim 0.2$.

The RMS variations of computed $A$ values using
different sky patches is almost always lower than the computed random errors 
for each 
asymmetry measurement. The average random asymmetry errors are plotted as a 
function of
redshift and magnitude in Figure~5 as solid lines, and are listed in 
Table~2.  The final corrected asymmetries values are plotted as a function
of redshift in Figure~5. 

\section{Methodology}

\subsection{The Merger Criterion}

A major goal in contemporary astrophysics is determining how galaxies
formed and evolved.  As we approach this goal, it is fair to ask if 
the structures of galaxies give any clues towards solving this problem.
Is it possible that the
morphological appearance of a galaxy is only representative of temporary
`weather', and does not relate to the fundamental underlying
evolution?  While we only briefly address these questions here, 
see Conselice (2003) for a detailed discussion and an introduction to
the CAS (concentration, asymmetry, clumpiness) classification system where
it is quantitatively argued that galaxy structures reveal fundamental
information.

In general, a galaxy's appearance or morphology is determined by a variety of 
different 
effects. Some of these are produced by projection, most notably in the case 
of edge-on spiral galaxies, but as argued in Conselice (2003) morphology is 
largely the result of physical processes, such as star formation, 
interactions and mergers with other galaxies, and the past history of
these events. 

As described in CBJ00 and Bershady, Jangren \& Conselice (2000), by using a
computationally consistent method to compute asymmetries there are
strong and physically meaningful relationships with other parameters, such 
as color and concentration (CBJ00).   This requires using a method for 
computing asymmetries that is not affected by choice of center, 
radius, or to first order, resolution. 

To determine how useful asymmetry and other parameters are for distinguishing 
galaxies in various phases of evolution, CBJ00, Conselice et al. (2000b) and 
Conselice (2003) computed asymmetries for over 200 nearby galaxies in all 
phases of evolution, including high-$z$ analogs such as starbursts and 
ULIRGs.   When asymmetries are combined with color, or other structural 
information, galaxy types such as ellipticals, spirals, and irregular 
galaxies can be roughly distinguished from each other.   As expected, 
ellipticals are symmetric, 
red objects; while later-types are both bluer and more asymmetric (CBJ00). 
We also argue in these papers that galaxies undergoing major mergers can be 
distinguished from those evolving quiescently through their global 
asymmetries. 

There are also fairly strong correlations
between (B$-$V) colors and clumpiness ($S$) values with the 
asymmetry index for non-mergers, such that more asymmetric galaxies are bluer 
and have higher clumpiness values.  
For these non-mergers there exists a small distribution of $A$ values 
at at all $S$ and (B$-$V) values, with a natural scatter $\sigma (A)$in
asymmetries (see Figures~7 \& 8 in Conselice 2003).   We define these 
distributions in Conselice (2003) and use them to identify statistical 
outliers with high asymmetries which we identify as mergers.
If we set a limit of  $A_{\rm merger}$ $>$ $A$($S$, B$-$V) 
+ 3$\sigma (A)$ for major mergers, then we find A(B)$_{\rm merger}$ 
$\sim$ 0.35 for both the bluest and most clumpy galaxies yet observed. 
Redder or less clumpy galaxies have a 3$\sigma (A)$ deviation less
than $A = 0.35$, thus we are being conservative with this limit.   From
a nearby sample of $\sim 240$ galaxies, nearly all objects that  deviate more
than 3$\sigma (A)$ from the asymmetry-color and asymmetry-clumpiness 
relationships (Conselice 2003) are galaxies involved in major mergers. 
As such, we use $A$(B)$_{\rm merger} = 0.35$ 
as our limit for identifying major mergers.

Other
evidence that $A$(B)$_{\rm merger} > 0.35$ includes galaxies 
with asymmetries larger than $A$(B)$_{\rm merger}$ showing kinematic evidence 
for merging based on 
broadened HI line profiles (Conselice et al. 2000b).  
The measured asymmetries of simulated galaxies in major merger simulations 
also have 
$A > 0.35$ when undergoing merging (Conselice \& Mihos 2003 in preparation).
These simulated galaxies have lower asymmetries with $A <$ 
$A$(B)$_{\rm merger}$ before, and after, merging events.

\subsection{Eddington Bias}

We argue in \S 4 that galaxies at higher redshifts are more asymmetric
in their rest-frame B-band morphologies.  We use this to further argue 
that, at the very least,
the most massive galaxies must be forming through the merger process.  
An important
question to ask however is how this result is biased by our observational
random errors, which in noisy data can mimic higher merger fractions, 
analogous to the aberration in star or galaxy counts due to
random errors (Eddington 1913).

The basic idea behind the Eddington bias is that within an intrinsic 
distribution of some observed quantity, in the presence of more and more 
noise, there will be larger measurement 
tails. In star counts the effect is to scatter more counts into bright bins 
from fainter bins.  In this section we investigate if the increase in 
$A$ values can be explained by increased noise scattering intrinsically low
asymmetry values into the high asymmetry bins.

To investigate the importance of Eddington bias in producing higher measured
asymmetries for galaxies at higher redshifts, we plot in Figure~6 the asymmetry
values for our sample divided into different redshift and absolute
magnitudes.  We also list in Table~2 the average random error of the
asymmetry values for the galaxies plotted in each of
the redshift/absolute-magnitude bins shown in Figure~6.  From these values,
and the other information in Figure~6, we argue 
that the Eddington bias is not a major effect when comparing 
asymmetries for galaxies at similar absolute magnitudes at 
different redshifts.  The errors remain below $\delta A = 0.10$
up until z = 3 when they become large.  

To show this, and to understand the limitations of our data and the effects of
random errors, we carry out Monte Carlo simulations, the results of
which are plotted on Figure~6, as the number in the lower right of each
panel.  This number is the sigma likelihood that an increase in
claimed asymmetries between two bins is not due to an increase in random
errors at higher redshifts. That is, it is the significance that any increase 
in asymmetries is real and not due to Eddington bias.

These simulations are done by assuming
that the random errors are Gaussian distributed, with a full width at
half maximum given by the increase in the error listed in Table~2.  To
determine how significant our observed increase in asymmetries are at
a given magnitude, we add
in resulting random errors to the asymmetries for galaxies in the
lowest redshift bin ($0 < z < 1$).  We
then recompute the mean and standard deviation of the resulting asymmetry
distribution.  This allows us to determine the likelihood
that an increase in average asymmetries, and
$1 \sigma$ distributions, at different redshifts are due to an increase
in random errors induced by being at higher redshifts.

The result of these simulations is that in almost
all cases, the statistical probability of a claimed asymmetry increase 
being due to
a random configuration of increased errors is very small.  Note
that we do not claim an increase between each redshift
interval for every magnitude (see \S 4.4).  From these
simulations we are confident at the 3 - 10 $\sigma$ level that all
claims for increases in
asymmetries between different redshifts is a real effect, and not a result of
Eddington bias.  The only exception is the change in asymmetries seen
for the $-20 >$ M$_{\rm B} > -21$ galaxies between $0 < z < 1$ and $1 < z < 2$.
This is the {\em only} interval in which a claimed increase in
asymmetries has a significant less than 3 $\sigma$.

Note that we do not use the $z > 3$ bin in our analysis
and only show it here for comparison purposes.
We latter quantify the Eddington bias in another way by always 
plotting the  $\pm$ random error merger fractions.  Doing this, we find
that the trend of increasing asymmetries at higher redshifts is 
still present, verifying the statistical argument made in this section.

\subsection{Contamination from Projection}

One potential problem with the asymmetry methodology is that occasionally
two galaxies will overlap to produce a two dimensional image that looks
asymmetric.  The two galaxies themselves can look very smooth and symmetric,
but when projected on or near each other they could
potentially be identified as a major merger through a resulting
high asymmetry.  Although we only measure asymmetries for galaxies
within well-defined radii produced through SExtractor, which deblends objects,
there are still possible cases of near neighbors that could produce a higher
signal.

As such, we must be able to constrain the importance of this effect.
Overlapping galaxies in the Hubble Deep Field are, however, rare, with only
a handful of obvious cases (White, Keel \& Conselice 2000).
Through a visual examination of all objects identified as a merger through
asymmetries, only seven are galaxies that might be in pairs (Figure~7).  
We say `might' here as it is often the case that these pairs
are real physical associations, and some of the photometric redshifts
suggest that they are in fact two nearby galaxies, perhaps in the early
phases of a major merger.  In any case, since our sample contains $\sim$ 
70 galaxies
identified as mergers, only 10\% of our mergers have something that could
resemble a pair, although many of these show evidence for being real physical
units due to similar colors, and in at least one case similar spectroscopic
redshifts, and thus are not chance projections.

\section{Results}

\subsection{Merger Candidates in the Hubble Deep Field}

All galaxies brighter than M$_{\rm B} = -20$ between $0 < z < 3$ found in the
HDF are shown in Figure~7 in the F814W band, divided by redshift 
intervals: $0 < z < 1$, $1 < z < 2$, and
$2 < z < 3$.  The mergers brighter than M$_{\rm B} = -20$ are also listed
in Table~3.   The images shown in Figure~7 are the appearance of each galaxy
in F814W, with three numbers overplotted, 
which are from top 
to bottom: absolute magnitude, rest-frame B-band asymmetry 
($A$(B)$_{\rm rest}$), and redshift.    Galaxies which are
statistically likely to be merger candidates, with $A$(B) $>$ 0.35, have
a solid box in the upper right corner of their panel.

This figure shows, among other things, that the asymmetry index is able to 
pick out systems that would be chosen as mergers through visual estimates. 
Systems with high asymmetries are however clearly in various phases of merging.
Some objects, particularly at high redshifts, 
appear to be two galaxies beginning to undergo a merger.  Some
also appear to have a low signal to noise morphology. 
This demonstrates 
the power of the asymmetry index to remove ambiguity and subjectivity in
determining which galaxies are merging.  While any given
handful of galaxy morphologists would pick out different mergers by eye
from this list, we are free from this concern as our method is purely 
automated, and can be understood statistically.  

As discussed in Conselice (2003), CAS parameters at high redshift are
best used in an ensemble sense, such as finding the fraction undergoing major 
mergers, while any one measurement can be dominated by random errors. 
For example, the asymmetry values listed on Figure~7 sometimes have 
random errors as large as 
$\delta A$ = 0.2, as do galaxies with $A(B) < 0.35$.  These errors can
remove, or add, galaxies into the merger bin, a fact that we account
for in \S 3.2 and \S 4.4 when analyzing the merger history of field galaxies.

\subsection{Physical Properties of Asymmetry and Size}

\subsubsection{Asymmetries}

In Figure~5a we plot the rest-frame asymmetries of our sample
galaxies, brighter than M$_{\rm B} < -18$, as a function of redshift
and luminosity.  In Figure~5b we plot the asymmetries as a function of
redshift and stellar mass for systems with M$_{*} > 10^{8}$ \solm.
The symbols on Figure~5 represent the magnitude or stellar mass of each
galaxy, with larger symbols representing brighter or more massive galaxies.   
Note that, particularly at high redshift ($z > 1$), 
some of the galaxies consistent with mergers are relatively
bright and massive.  

The average asymmetries of objects brighter than the given magnitude limit  
listed in Table~4 are plotted in Figure~8 as a function of redshift, where the
respective bright magnitude limits are labeled next to each line.  
The average asymmetries generally increase from the $z = 0 - 1$ range to
the $z = 1 - 2$ range, and then decreases at 
higher redshifts.  This is especially true at the fainter magnitude
limits.  At brighter magnitudes the average asymmetry generally 
increases with redshift, peaking at $z \sim 2.5$, and declining
thereafter.  At the brightest magnitude limit of M$_{\rm B} < -22$ the 
average asymmetry increases greatly between $z \sim 1$ and 1.5 and then 
stays large at higher redshifts.

\subsubsection{Colors}

We can get some idea of the stellar populations that make up galaxies seen
in the HDF by examining their rest-frame Johnson
(B$-$V) colors.    These colors are derived from an interpolation of
the broad-band photometry, based on the known spectroscopic redshift, or
computed photometric redshift, for each object.  A plot of rest-frame color 
versus redshift is shown in
Figure~9 out to $z \sim 2.5$, the limit where we can measure
rest-frame (B$-$V) colors.  An obvious feature of this plot is the presence of
a large population of faint and very blue galaxies at redshift above
$z \sim 1.5$ which are not seen at lower redshifts.  Some of these
galaxies are systems with low asymmetries seen in the corresponding diagram of
asymmetry versus redshift. 

Many of these very blue and faint galaxies are 
however the result of either noisy photometry, imprecise photo-zs, bad 
K-corrections,
or a combination of these effects, as very few nearby galaxies have negative 
(B$-$V)
colors.  Population synthesis models also show that objects with mean weighted
single stellar population burst ages of 50 Myrs, with no subsequent 
star formation, have a color (B$-$V) $\sim 0.04$ (Bertelli et al. 
1994).  In fact, since most of these colors are derived from spectral
energy distribution fits to broad brand photometry, there are likely to
be some systematic errors in individual colors.  We therefore only use
these colors in later figures as representative ensembles in different
populations.   The values of these colors also do not affect any of
the quantitative results of this paper.
There does appear, based on Figure~9, to be an absence of bright red galaxies 
with
M$_{\rm B} < -20$ and (B$-$V) $> 0.5$ at $z > 1.5$, although we cannot
out rule that some intrinsicly red galaxies appear blue on Figure~9 due
to systematic effects.

\subsubsection{Asymmetry and Color Evolution in the Bright Galaxy Population}

The relationship between asymmetry, color and magnitude for HDF galaxies is
shown by plotting rest-frame B-band asymmetries as a function of M$_{\rm B}$ 
into two different redshift bins in Figure~10.  Figure~10a and Figure~10b show 
M$_{\rm B}$ vs. 
A(B) where the points are colored according to their rest-frame
(B$-$V) values, within 
the redshift ranges 0 $ < z < 1.5$ (Figure~10a) and 1.5 $ < z < 3$ 
(Figure~10b), respectively.  This roughly divides the sample into
galaxies dominated by visible star-formation and those that are not. 
Figure~10a shows that the brightest galaxies in the low-redshift bin
have low asymmetries and red (B$-$V) colors, consistent with smooth galaxies 
with old stellar populations, such as ellipticals.  There is also a 
clear bifurcation in Figure~10a such that
asymmetric galaxies are generally blue, and the low-asymmetry objects
are generally red (see \S 4.3), the bluer objects possibly the result of merger
induced starbursts. 
We see the opposite however at higher redshifts, $1.5 < z < 3$, (Figure~10b) 
where the
brightest galaxies tend to be blue objects undergoing starbursts,  which
have high asymmetries, suggesting they are potentially major mergers.

\subsubsection{Size Evolution}

Another property that we investigate is the size evolution of
galaxies in the HDF.  For this we use the 0.5$\times k_{r} \times a$ radius
within which asymmetries are measured.  Like all the other parameters 
discussed in this paper, the apparent sizes of objects depends on redshift,
as dictated by the angular size distance.  However, the sizes of objects
may also appear smaller due to surface brightness dimming that makes
the outer parts of galaxies difficult to detect.  To address this issue
we investigate, through the simulations discussed in \S 2.3.5, how measured
sizes of galaxies change due to cosmological surface brightness dimming.  
The change in galaxy sizes due to this effect are
plotted in Figure~4b, where average radius differences and their
1$\sigma$ variations are plotted as a function of redshift.  

When we apply this correction to the measured sizes in the rest frame
B-band, and examine the 
resulting distribution as a function of redshift, we get Figure~11.   There
are two interesting properties in this figure. The first is that
there appears to be a lower limit on galaxy sizes, which is
around 3 kpc. This
limit is likely the result of the SExtractor detection
method and effects from the PSF, as galaxies smaller than this certainly 
exist in the local universe (e.g., Conselice et al. 2002).   
There is in fact a bias in the way
that SExtractor measures the sizes of galaxies, which typically depends upon
luminosity. Therefore, the best way to view Figure~11 is as
a relative change in galaxy sizes with redshift.
The range and distribution of galaxy sizes appears roughly constant
between 1.4<z<3, while for z<1.4, there are are significantly more
galaxies with radii above 7 kpc.
Short of a significant amount of cosmic 
variance, this evolution is real, as these sizes have been corrected for
cosmological dimming.  There therefore appears to be
no galaxies at $z > 1.4$ larger than 10 kpc in the HDF-N, even after
correcting for redshift effects. 

This is consistent 
with the idea, but does not prove, that nearby large galaxies are forming 
from the mergers of
lower-mass, and presumably smaller, galaxies.  An alternative interpretation
is that these galaxies are still forming (inside out) from accreted 
intergalactic gas cooling into stars
after $z \sim 1.5$.  Size and luminosity evolution (Ellis et al. 1996) 
alone do not prove that mergers are occurring in the galaxy population, but 
both effects are consistent with this idea.  
For the remainder of the paper we examine the structures
of galaxies in the Hubble Deep Field and argue that
mergers are increasingly common out to $z \sim 3$, especially for the most
massive systems.

\subsection{Star-Formation versus Merger Formation}

We can do a general test to determine if, and approximately 
how much, star formation is 
the cause of asymmetries in HDF galaxies by examining their star forming
properties as traced by color.
If star-formation is responsible 
for producing asymmetries of our sample, then we would
expect more asymmetric galaxies to be dominated by star formation.

We can use the color-asymmetry diagram to investigate this question.  
The color-asymmetry diagram, as introduced in
Conselice (1997), and discussed in CBJ00 and Conselice et al. (2000b) is a 
diagnostic tool that plots the disturbance of a galaxy with a 
measure of its spectral shape, which signifies the ages of its stellar 
populations.  Figure~12 shows the HDF rest-frame (B$-$V) color-rest frame 
B-band asymmetry A(B$_{\rm rest}$) diagram plotted in bins of absolute 
magnitude, M$_{\rm B}$.  The differences and
similarities between the local galaxy population (see CBJ00) and
the HDF galaxies can be seen.  The diagonal line is the 
relationship between asymmetry and color 
characterized by CBJ00 for nearby normal galaxies. The solid part is an 
extrapolation of this relationship to colors bluer than (B$-$V) $\sim 0.4$, 
which is typically among the bluest colors found for e.g., dwarf irregulars in
the nearby universe.   While there are
many galaxies with modest asymmetries and blue colors, there are few analogs
of nearby ellipticals in the HDF
with red colors and low asymmetries (CBJ00). 

Figure~12 can be used to argue that high asymmetries are not likely
solely produced by massive amounts of star formation.  At each magnitude
and redshift range there exist a range of $A_{\rm rest}$ values for
any given color. In fact, the galaxies consistent with mergers (those to
the right of the vertical dashed line) do not have colors that significantly
differ from blue non-mergers.  They are, on average, bluer than the total 
population, but there are always galaxies with similar, or bluer,
colors that do not have similarly high asymmetries (this can also
been seen in Figure~10).  

If vast amounts of star formation are 
responsible for large asymmetries, then the galaxies with the bluest
colors should have the largest asymmetries, which is clearly not the case.
The galaxies with high asymmetries are unique in terms of
their structure, which we interpret for the reasons in \S 3 and 
Conselice (2003) that they are undergoing major mergers.

\subsubsection{Galaxy Populations in the Color-Asymmetry Diagram}

As mentioned earlier, there appears to be a lack of corresponding
modern day ellipticals, or 
very red-symmetric objects at any redshift bin in the HDF.  This 
may simply be due to the fact that ellipticals are bluer at high
redshift and/or due to the very small co-moving volume of the
HDF at low redshift which is 320 h$^{-3}$ Mpc$^{3}$ at $0.1<z<0.5$, where
few, if any, galaxies with L$>$L$_{*}$ are present (Dickinson et al. 2003).  
Morphologically selected elliptical-like galaxies however do exist in the
HDF, particularly around $z = 1$ (Stanford et al. 2003).  
They are however bluer than nearby
ellipticals as their stellar populations are younger then those
in nearby ellipticals (Menanteau et al. 1999), an effect also seen in high
redshift clusters (e.g. Dickinson 1997; Stanford, Eisenhardt \& Dickinson 
1998).   Passive evolution from $z \sim 1$ to $0$ will create a reddening
of about 0.1 magnitudes, which would place these symmetric bluer galaxies in
the general area of nearby ellipticals (CBJ00).

Figure~12 further shows that many galaxies in the HDF 
are potentially undergoing mergers with asymmetries $A$(B)$ > 0.35$.  These 
objects have asymmetries inconsistent with being normal, late-type galaxies 
undergoing quiescent star formation as compared to nearby galaxies.
As we go to higher redshifts the fraction of galaxies with asymmetries
consistent with merging increases (Figure~12; \S 4.4). 

What are the galaxies in the HDF that are not identified as mergers?  At low
redshifts, these are simply the morphologically familiar disks and 
ellipticals (Figure~7).
At higher redshifts, the galaxies inconsistent with merging are not
readily identifiable, through eye-ball estimates, with any specific local 
morphological type, but
span a range, from very compact galaxies with possible 'tidal' features to
diffuse blob like systems (Figure~7) (see also Giavalisco, Steidel \&
Macchetto 1996).  There are also symmetric systems at redshifts $z > 1.5$,
although these are generally blue. 

\subsection{Galaxy Merger Fractions and their Evolution}

In this section we use the asymmetries of HDF galaxies to measure the 
evolution of 
implied major merger fractions out to $z \sim 3$.
As in the previous sections we avoid morphological K-corrections by using the 
rest-frame B-band asymmetries of galaxies in the HDF and applying the
corrections as needed at higher redshifts as described in \S 2.3.5.

As previously described, we define a major merger as a galaxy whose rest-frame 
B-band asymmetry is larger than $A_{\rm merger} = 0.35$ for the reasons 
discussed in \S 3. By taking the 
ratio of galaxies with asymmetries $A > A_{\rm merger}$ to
the total number of galaxies in a given parameter range, the implied merger 
fractions out to $z \sim 3$ can be computed as a function of
absolute magnitude M$_{\rm B}$, stellar mass (M$_{\star}$), and 
redshift ($z$).  
This allows us to determine how different galaxy types have evolved as a 
function of mass and time.  
These merger fractions are listed and plotted in four different lower 
galaxy absolute magnitude limits, 
M$_{\rm B} = -18, -19, -20, -21$ in Table~5 and Figure~13, and mass limits 
M$_{\star}$ = 10$^{8}$ \solm, 10$^{9}$ \solm, 10$^{9.5}$ \solm, and
10$^{10}$ \solm in Table~6 and Figure~14.  Each merger fraction is
computed for galaxies brighter, or more massive, than these limits.  
For example, the M$_{\rm B} = -18$
bin contains all the galaxies in the M$_{\rm B} = -19, -20$ and $-21$ bins.

The solid large circles in Figures~13 and 14 are the inferred merger fractions
based on the asymmetry measurements using the various magnitude and mass
limits shown on the panels in Figures~13 and 14.  
The high and low valued green crosses
are the computed merger fractions after adding and subtracting, respectively,
the 1$\sigma$ error from each asymmetry measurement, and then recalculating the
merger fractions.   This is a graphical representation
of a likely outcome of the effect of random errors and systematic biases
on the measured distribution of asymmetries (i.e., the Eddington bias) 
(\S 3.2).

Merger fractions from the magnitude
selected pair studies of Patton et al. (1997) and Le~F\'{e}vre et al. (2000) 
and the kinematic pairs study of Carlberg et al. (2000) are also 
plotted on Figure~13 at their most representative magnitude regime.
These authors compute merger fractions by finding the number of galaxies
at each redshift range separated by some projected
distance (usually $< 20$ kpc), and in the case of Carlberg et al. (2000), a 
relative velocity difference ($< 500$ \kms).  These authors however use various
magnitude limits, and it is straight forward to compare each of these points 
to our
values.  Carlberg et al. (2000) use a magnitude limit of M$_{\rm B} \sim -20.5$
while Patton et al. (1997) and LeFevre et al. (2000) find pairs within
magnitude limits of M$_{\rm B} = -18$ and M$_{\rm B} = -19$, respectively.
None of these studies use limiting stellar masses as we do here.  

Note that we may be missing a substantial number of dusty merging galaxies, 
such as sub-mm sources, 
which we are not considering because they are too faint to be detected within 
our
limits.  These would increase the merger fractions above those calculated 
here for galaxies within our stellar mass limits 
(Burgarella et al. 2003).  They would not however effect the derived
merger fractions for systems within the given magnitude limits.

\subsubsection{Fitting Functions}

We characterize the evolution of these merger fractions by fitting a 
simple power-law increase with redshift:  
f$_{m}$($A$, M$_{\star}$, M$_{\rm B}$, $z$) = f$_{0} \times (1+z)^{m_{A}}$,
where f$_{0}$ is the merger fraction at $z = 0$ and
$m_{A}$ is the slope of the merger fraction evolution, such that more
steeply rising merger fractions have higher $m_{A}$ values.  We perform
these fits to quantitatively parameterize how merger fractions evolve, and
to compare with previous work and theoretical studies that also use 
this parameterization.
We have fit this model to the data in two different ways: a simple unweighted
least-squares fit to all the merger fractions and by fitting the merger
fractions by holding the $z \sim 0$ point to the value found
by Patton et al. (1997).  

The results of these various fits are plotted as two different colored lines
in Figures~13 and 14, with the fitted parameters
listed in Tables~7 \& 8. 
A fit to only the asymmetry merger fractions are shown as the straight dark 
blue
lines on Figures~13 and 14.  The $m_{A}$ values for these fits are fairly
low for the fainter M$_{\rm B} > -20$ (Table~7) and low stellar mass systems
with M$_{\star} <$ 10$^{9.5}$ \solm (Table~8) with
typical $m_{A}$ values $\sim$ 0.5 - 1.  
However we find that for the brightest and highest mass systems with 
M$_{\rm B} < -21$ and M$_{\star} >$ 10$^{10}$ \solm the merger slopes are 
quite steep with values of 3.7$\pm$0.3 and 5.9$\pm$1.3, respectively.  
As we argued in \S 2 and \S 3.2 this increase in asymmetry between
different redshifts cannot be accounted
for by morphological k-corrections, detection incompleteness, systematic
errors, random errors or Eddington bias.

If we use the Patton et al. (1997) merger results as a fiducial $z \sim 0$ 
bench-march for all magnitude and mass cuts, and then fit the
merger fraction evolution based on this, we get higher fitted 
slopes
with values near $m_{A} \sim 1.5 - 2$.  These fits are shown in Figures~13
and 14 as the cyan colored lines.  

We can also examine merger fraction evolution when only
fitting up to certain redshifts.  Tables~7 and 8 list the values of 
$m_{A}$ for these fits out to $z \sim 1$ and $z \sim 2$ as a function of 
limiting stellar mass and magnitude.  At these lower redshift limits
the merger fraction slope, $m_{A}$, becomes quite steep, except for
the brightest and most massive systems (Tables~7 \& 8).  The slopes of
these fits are typically $m_{A} = 2.5 - 3$ for galaxies with 
M$_{\rm B} > -20$ or
M$_{\star} <$ 10$^{9.5}$~\solm.  At $z \sim 2$ the merger fractions are lower
and the fitted slopes, $m_{A}$, decrease to $m_{A} \sim 2$ for systems with 
M$_{\rm B} > -20$ or
M$_{\star} <$ 10$^{9.5}$~\solm.  We see the opposite effect however
for the most massive systems, with  
M$_{\star} >$ 10$^{10}$~\solm.  For these galaxies the fitted power-law
slope, $m_{A}$, is quite
low, between $z \sim 0$ to 1, with a value of $m_{A}$ = 1.7.  
The brighter systems with M$_{\rm B} < -21$ also have lower $m_{A}$
slopes between $z \sim 0 - 1$ with values $m_{A} \sim 1.5$. 

This bifurcation in $m_{A}$ values (Tables~7 \& 8) between
bright/massive and faint/lower-mass galaxies suggests that massive
galaxies form
from mergers much earlier than lower mass systems. The steep 
decline in merger fractions suggests the most massive galaxies
underwent a massive merger phase in the distant past
and have become the quiescent massive galaxies (or quiescent massive galaxy
components) we see in the nearby universe.

\subsection{Merger and Stellar Mass Assembly Rates}

Using further assumptions
we can investigate the merger and mass accretion rates of galaxies from $z = 0$
out to $z \sim 3$, or back to when the universe was only $\sim$ 2.1 Gyrs old.
It should be kept in mind however that the calculations we perform
below are somewhat speculative as we do not yet have a firm understanding
of many of the properties we are assuming throughout the following analysis.
One possible problem with this is that the Hubble Deep
Field occupies only a small volume of space and a large cosmic
variance may make these results inapplicable globally.  
We know that in dense areas such as clusters, the merging properties
are probably different than in the field (e.g., van Dokkum et al. 1999).
Despite this, the Hubble Deep Field samples the high-redshift
universe and thus within the limitations imposed by the uncertainties in our
assumptions, we can compute certain evolutionary quantities for the first time.

\subsubsection{Merger Rates}

To understand the merger and mass accretion rates we must know the dynamical 
time-scale, and mass ratios, of mergers which will result in systems
with high asymmetries.  
That is, we have to have some idea
of what type of major merger will produce a galaxy structure with an
asymmetry larger than $A_{\rm merger}$. The two most important parameters
for understanding this are:
the time scale which a galaxy will remain asymmetric, such
that $A > A_{\rm merger}$, and the
mass ratios necessary to produce $A > A_{\rm merger}$.  
The time-scale for a merger also varies with redshift solely due to
the different physical conditions of galaxies in the past, although
we ignore this for the moment and assume that mergers occur within
the same time-scale at all redshifts.

Estimates of mass merger ratios and time scales can be computed through
N-body simulations of galaxies undergoing the merging process.
These simulations have been done and are fully reported in Conselice
\& Mihos (2003 in preparation; hereafter CM03).  We briefly summarize the 
conclusions of 
this study and use this information to determine the merger and mass assembly 
rates for galaxies seen in the HDF.  

The merger of two
galaxies of nearly equal mass remains asymmetric for roughly 0.9 Gyrs,
that is $A > 0.35$ during this length of time (CM03).
There is however a variation in this time range, partially for
mergers between systems that do not contain similar masses.  Also,
we have no a priori method of determining the mass ratios of the systems that
produced the galaxies which we see as mergers.  To simplify this,
we assume that each merger is produced from two galaxies 
of approximately equal stellar masses.

Using this modeled length of time as the average time scale for which
$A > A_{\rm merger}$, we can compute the merger rate as a function 
of redshift.   We measure this by assuming, as earlier, that each
galaxy with $A_{\rm B}$ $>$ $A_{\rm merger}$ is currently undergoing a major 
merger 
that lasts for $\sim 0.9$ Gyr, within the co-moving volume between each 
redshift interval.  Using these number, we can then calculate the merger
rate $\dot{\phi}_{M}$, defined as the number of mergers occurring per co-moving
volume (in Gpc$^{3}$), divided by the time-scale of the merger (0.9 Gyrs).

These merger rates, $\dot{\phi}_{M}$, for galaxies
brighter than M$_{\rm B} = -19$ are plotted as a function of 
redshift ($z$) in Figure~15 as a solid line.  We calculate the
merger rate as 
$\dot{\phi}_{M}$ = 3.6 $\times$ 10$^{5}$ Mergers Gpc$^{-3}$ Gyr$^{-1}$ 
between $z \sim 0.4 - 0.8$ and find a peak rate of  $\dot{\phi}_{M}$ = 
2 $\times$ 10$^{6}$ Mergers Gpc$^{-3}$ Gyr$^{-1}$ between $z = 0.8$ and 1.4.
At redshifts  higher than $z \sim 1$ the merger rate declines towards the 
merger rate values found at $z < 1$.  These rates, as well as the 
mass accretion rates (\S 4.5.2), are listed in Table~9.  From Figure~15
the merger rate appears to be nearly constant, but the merger fraction 
generally increases with redshift (Figure~13-14).  This difference is due
to the fact that we are sampling more volume at higher redshifts, and thus
the number of mergers per unit volume remains relatively flat.

\subsubsection{Mass Accretion Rates}

The stellar mass accretion rate per co-moving volume ($\dot{\rm M}$) is 
plotted as the dashed line and on the right axis of Figure~15.   The stellar
mass accretion rate is computed in a similar way to the merger rate, 
$\dot{\phi}_{M}$.  This is found by assuming that each major 
merger with $A > A_{\rm merger}$ consisted previously of two galaxies of 
equal mass that merged.  Therefore the mass accretion rate, $\dot{\rm M}$, is 
the sum of half the total mass involved in mergers in a redshift
interval divided by the co-moving volume in that interval, divided by the 
major merger time scale.
The resulting $\dot{\rm M}$, roughly follows the form of the merger
rate with a peak value of 6.0$\times 10^{6}$ \solm Gpc$^{-3}$ year$^{-1}$
between $z = 0.8$ and 1.4.  

It may seem surprising
that the mass accretion rate and merger rate track each other very well
out to $z \sim 2.5$ since the masses of high redshift galaxies are
on average smaller than those at low redshift (Dickinson et al. 2003).  If
galaxies involved in mergers were drawn from a random distribution, then
for a merger rate which is constant with $z$ would correspond to a
mass accretion rate that declines with $z$.   However, because the
galaxies involved in mergers are of higher mass at higher redshifts, the
mass accretion rate does not decline.

The stellar mass accretion rate
per galaxy $\dot{\rm M}_{\rm G}$ is also plotted as a function of redshift in 
Figure~16.  The values of  $\dot{\rm M}_{\rm G}$ are calculated by determining
the amount of stellar mass added to galaxies from major mergers in a redshift
interval divided by the total number of galaxies within that interval per
merger time scale (0.9 Gyr).  It is the average stellar mass accreted onto a 
galaxy due to major mergers per unit time. 

The stellar mass accretion rate per galaxy, 
$\dot{\rm M}_{\rm G}$, for galaxies more massive than 10$^{9}$, 10$^{9.5}$ 
and 10$^{10}$ \solm are plotted on Figure~16. For galaxies
at M$_{\star} >$ 10$^{9}$ the peak rate is 
5.5$\times 10^{8}$ \solm Galaxy$^{-1}$ Gyr$^{-1}$ at $z \sim 2.5$.  The
peak rate is 1.16 $\times 10^{9}$ \solm Galaxy$^{-1}$ Gyr$^{-1}$ and
4.72 $\times 10^{9}$ \solm Galaxy$^{-1}$ Gyr$^{-1}$ for systems
at M$_{\star} >$ 10$^{9.5}$ \solm and M$_{\star} >$ 10$^{10}$ \solm, 
respectively.
The evolution of $\dot{\rm M}_{\rm G}$ can be further parameterized as: 

\begin{equation}
\dot{\rm M}_{\rm G}(z) = 1.6 \times 10^{8}~\solmm {\rm Galaxy}^{-1} {\rm Gyr}^{-1} (1+z)^{0.99\pm0.32} 
\end{equation}

\noindent for systems with M$_{\star} >$ 10$^{9}$ to $z = 2.5$. 
The best fit for systems at M$_{\star} >$ 10$^{9.5}$ is

\begin{equation}
\dot{\rm M}_{\rm G}(z) = 1.8 \times 10^{8}~\solmm {\rm Galaxy}^{-1} {\rm Gyr}^{-1}(1+z)^{1.47 \pm 0.25} {\rm and}
\end{equation}

\begin{equation}
 \dot{\rm M}_{\rm G}(z) = 5.3\times 10^{6}~\solmm {\rm Galaxy}^{-1} {\rm Gyr}^{-1} (1+z)^{5.3 \pm 0.16}
\end{equation}

\noindent for galaxies with
M$_{\star} >$ 10$^{10}$. These best fits are plotted as dashed lines on 
Figure~16.

\subsubsection{Building Galaxies Through Mergers}

Using the derived values in \S 4.5.2 we can investigate how much of
the stellar mass of modern galaxies is formed through major mergers.
This can be done by integrating equations 1-3 from $z \sim 3$ to $z \sim 0$
where we observe galaxies in their  most evolved form.  
By integrating equations 1-3 we find that
the addition of mass for these galaxies in the Hubble Deep Field North
due to mergers is at most $\sim 10^{10}$ \solm since $z \sim 3$.  This
shows that on average the most massive galaxies only slightly
double in stellar mass due to major mergers.  This is not enough stellar 
mass to produce an $L^{*}$ galaxy at $z \sim 0$, which has a stellar
mass of $\sim 10^{11}$ \solm.  However, the vast amounts of ongoing
star formation in these systems, probably induced by this merging, will
form stars along with this addition of already existing stellar mass obtained
through major mergers.  The HDF may also be devoid of the most massive,
$> L^{*}$, galaxies, as more massive systems at $z > 2$ do exist (Shapley
et al. 2001).  This is certainly a volume effect and these massive systems
are rare, and the more typical galaxies found in the HDF are those
that dominate the mass density.

Equations (1-3) reveal the first mapping of galaxy stellar mass assembly from
mergers as a function of stellar mass.  Combining this with star formation
histories, and the resulting stellar mass formation histories
(Dickinson et al. 2003),
it is possible to determine the formation histories of galaxies from
$z \sim 3$ as a function of their initial stellar mass.

\subsubsection{Mass and Luminosity in Mergers}

We use the above information to determine the fraction of stellar 
mass, and rest-frame B-band luminosity, involved in major 
mergers, within our stellar mass and magnitude limits, as a function of
redshift.   These fractions are
calculated by determining the total amount of mass and light in all
galaxies within a given
redshift interval, and comparing this to the amount of light and mass
within that interval that are attached to galaxies involved in major
mergers.

Figure~17 shows that roughly 5\% to 25\% of the stellar mass and luminosity 
in HDF galaxies are involved in major mergers
for systems with M$_{\rm B} > -20$ or M$_{\star} < 10^{9.5}$ \solm.  
For the highest mass and brightest systems
with M$_{\star} > 10^{10}$ \solm or M$_{\rm B} > -21$ there is a high mass 
fraction peak of 0.5, and luminosity fraction peak of $\sim 0.6$, at
$z \sim 2.5$ and a rapid decline at lower redshifts.  

This again demonstrates that
high mass and luminous galaxies underwent major mergers at high redshift, but
are not doing so in the nearby universe.  Figure~17 also shows that
although the fraction of stellar mass in the brightest and most massive 
systems involved in major mergers declines at lower redshift, the fraction of
the luminosity coming from major mergers is relatively constant until
$z \sim 1$, when it drops.  It should be kept in mind
however, that the stellar masses used to create Figure~17 are from the
stars that dominate the light of these galaxies and there is a possibility
that these galaxies are hiding a significant amount of mass in
the form of old stars (e.g., Papovich 2002; Dickinson et al. 2003).
These results are however consistent with these massive and bright 
galaxies being ellipticals, or bulges formed through
mergers $\sim 9$ Gyrs ago.  It is also consistent with peculiar galaxies
at higher redshift transforming into normal Hubble types (e.g.,
Driver et al. 1998; Brinchmann \& Ellis 2000). 

\subsubsection{Possible Biases}

The evolution in Figures~16 and 17 appears steepest for
objects above the largest mass threshold, i.e. $> 10^{10}$ \solm.  Given
the stellar mass evolution inferred for the general galaxy
population from Papovich et al. (2001) and Dickinson et al. (2003), it
is worth noting that a fixed mass threshold such as $>$ 10$^{10}$ \solm 
corresponds to increasingly
rare and unusual objects at higher redshifts.  For example, at $z \sim 0$, 
10$^{10}$ \solm is $\sim 0.1$ of the mass of an L* galaxy, while at $z \sim 3$
it appears to be more or less the M* mass of Lyman-break galaxies,
that is more massive objects are
evidently rare.  Thus, by using a fixed mass threshold, we are
examining rates for objects that are increasingly extreme at higher
redshift, and which (in hierarchical clustering theory) are themselves
ever more strongly biased (in terms of their clustering) relative to the
underlying dark matter density distribution.

This might have some implications for the results we see 
in Figures~16 and 17.  The steepness of the apparent evolution at high 
masses might be an artifact due to this effect, although this is not
likely the case.  Unlike the mass function, the rest-frame optical luminosity
function does not evolve dramatically with redshift, and yet there still
is a rapid evolution for objects with M$_{\rm B} < -21$ (Figure 16).  These
bright objects, unlike galaxies with M$_{\star} > 10^{10}$ \solm, are not 
increasingly rare at higher redshift.

\subsection{Comparison to Models}

Many papers have recently made merger history predictions based on the CDM 
paradigm (Khochfar \& Burkert 2001; Gottlober, Klypin \& Kravtsov 2001)  
which we compare to our results.
The idea that galaxies build up from mergers began with the 
modeling of this process by Press \& Schechter (1974)
and White \& Ress (1978).  These early models traced the abundances of
dark halos as a function of time with some assumptions for how galaxies
form through gas cooling and feedback. 
Later, these models were expanded to include an initial power-spectrum of
dark halos assuming that dark matter is cold (White \& Frenk 1991).  
This Cold Dark Matter (CDM) approach using an extended Press-Schechter
formalism (Bower 1991; Bond et al. 1991) has
proven to be successful at reproducing many nearby and high-redshift 
galaxy properties, through semi-analytic and N-body
modeling (e.g., Kauffmann, White
\& Guiderdoni 1993; Cole et al. 1994; Guiderdoni et al. 1998; 
Kauffmann et al. 1999; Sommerville 
\& Primack 1999; Cole et al. 2000; Somerville, Primack \& Faber 2001).
However, the basic
idea that galaxies and dark halos merge, the fundamental idea
of CDM models, has until now never been tested observationally at high
redshifts where most galaxy formation occurs. 

There are several results from this paper which we can quantitatively 
compare with CDM predictions of
structure formation.  A basic test is to examine how merger 
fractions
change as a function of redshift in CDM models, within magnitude and stellar 
mass limits.  Comparisons with the semi-analytic model results of Benson et 
al. (2002) are shown in Figures~13 and 14 as solid red lines.  These 
predicted merger fractions were
computed by finding all major mergers,
defined, for purposes of this model, as systems merging with a mass ratio of 
1:3 or higher and within the
past Gyr, at a given redshift, in the semi-analytic simulation of Benson
et al. (2002).  
These are, as best as we presently know, the 
physical parameters within which the asymmetry index is sensitive.

These CDM merger fraction
predictions, agree fairly well with our computed merger fractions
at the highest redshifts. At the
highest masses (M $> 10^{10}$ \solm) and brightest magnitudes 
(M$_{\rm B} < -21$)
the difference between the observed merger fractions are within 1$\sigma$ of
the Benson et al. (2002) model results at $z > 2$.  Discrepancy 
is however present at some redshifts, and at various stellar mass and 
magnitude limits, particularly for the faintest and lowest mass systems.    
This discrepancy is highest at $z \sim 1$ for systems with low masses 
(M$_{\star} < 10^{9.5}$ \solm) and faint
magnitudes (M$_{\rm B} > -20$), where we find differences at significances of 
$ > 3 - 4  \sigma$.  At other redshifts 
we find that the differences are lower with significances of 
$\sim 0 - 2 \sigma$.   These differences result from the fact that in 
hierarchical
formation models, low-mass objects form by merging first, while it appears 
from our
observations that lower mass galaxies tend to form continuously throughout
the history of the universe, with a significant fraction of low-mass 
systems not undergoing mergers at any redshift.

We can also use merger fraction predictions from Khochfar \& Burkert (2001) 
to test qualitatively
how well CDM theory predicts observed merger fractions as a function of 
redshift,
stellar mass, and absolute magnitude.  Khochfar \& Burkert (2001) suggest that 
at brighter
limits the merger fraction power law fits have changing f$_{0}$ and
$m_{\rm A}$ values such that at higher mass and brighter 
magnitude limits f$_{0}$
goes down while $m_{\rm A}$ goes up.  In other words, the present-day merger 
fraction (f$_{0}$) is lower for more massive galaxies,
while the redshift evolution of the merger rate ($m_{\rm A}$) is steeper.  
We see this from 
the fits to the data (Tables~7 \& 8) and also in terms of the mass evolution 
of the galaxies we study 
(\S 4.4).  For the most massive and brightest systems we find that
f$_{0} \sim 0$ and
m$_{\rm A} \sim 4 - 6$, consistent with the prediction that these massive 
systems formed by merging early (Khochfar \& Burkert 2001).

The most interesting discrepancy between the computed merger fractions and
those predicted by CDM is at a redshift of $z \sim 1$.  Although some of
the significance in the difference ($\sim 4 \sigma$) can probably be 
accounted for by cosmic
variance, we speculate that it is unlikely to account for such a high
$z \sim 1$ galaxy merger fraction.  
While it is possible that there exists a significant number
of very low-surface brightness galaxies at $z \sim 1$ which we are 
missing, we are not likely missing normal galaxies at
these redshifts, as these are clearly seen when lower redshift galaxies
are placed at $z \sim 1 - 2.5$ (see \S 2.3.5).   We therefore conclude that
at $z \sim 1$ a physical effect is occurring which increases
the merger fraction, or we are missing a population of galaxies
for some unknown reason.  The fact that CDM simulations, which
have ad hoc prescriptions for creating star formation, and our merger
fractions based on a small area of the sky, agree so well is suggestive
that at least the most massive galaxies do indeed form by mergers.

\section{Summary}

In this paper we present the first direct evidence for the hierarchical 
assembly of massive galaxies, as well as  measurements of merger and
merger mass
fractions and rates at redshifts $z > 1$.  We are able to measure this
evolution by identifying galaxies undergoing major mergers out to $z \sim 3$
using the asymmetry parameter as described in Conselice et al. (2000a) which
is part of the CAS morphological system (Conselice 2003).
After simulating how our asymmetry measurements change due to
decreased S/N and reduced resolution inherent at higher redshifts, 
investigating completeness, the Eddington bias, and 
correcting for these effects, we are able to conclude the following:

\noindent i) The merger history of field galaxies changes as a 
function of absolute rest-frame magnitude M$_{\rm B}$, stellar mass 
(M$_{\star}$)
and redshift ($z$).  For galaxies with M$_{\rm B} > -20$ or 
M$_{\star} < 10^{9.5}$
\solm the merger fraction peaks at a value $\sim 0.2$ near $z \sim 1$
and slightly declines at higher redshift.  Fitting these lower mass
and fainter merger fractions to a simple power-law of the form 
f = f$_{0} \times (1+z)^{m_{\rm A}}$ we find power-law
slopes, $m_{\rm A} \sim 2.5-3$ out to $z \sim 1$ and $m_{\rm A} \sim 0.5 - 1$ 
from $z \sim 0 - 3$.    The corrected sizes of galaxies also become larger
at lower redshifts (Figure~11) (see also Papovich et al. 2003).

\noindent ii) We see a clear bifurcation in the merger fraction evolution for
the most massive, M$_{\star} > 10^{10}$
\solm, and brightest, M$_{\rm B} < -21$, galaxies such that the merger fraction
continues to increase for these systems  at higher redshifts, with peak 
values near 0.5 at
$z \sim 2.5$ and low fractions $\sim 0$ at $z \sim 0$.  Fits to the
power-law evolution of the merger fractions for these galaxies reveals very
steep slopes with $m_{A} \sim 4 - 6$.  That is, 
luminous systems underwent more frequent major mergers at high redshifts 
than lower-luminosity galaxies, or their low-redshift bright and massive
counterparts.

\noindent iii)  By using results from N-body simulations of galaxies involved
in major mergers we are able to convert merger fractions into co-moving
volume merger rates, $\dot{\phi}_{M}$,  finding a peak merger rate at 
$z \sim 1$ of $\dot{\phi}_{M} \sim 2 \times 10^{6}$ mergers Gpc$^{-3}$ 
Gyr$^{-1}$ for systems with M$_{\rm B} < -19$.

\noindent iv) Using stellar masses (M$_{\star}$) of the HDF galaxies measured
by Papovich (2002) we determined the evolution in the
mass accretion rate per co-moving volume, the mass accretion rate per
galaxy, and the fraction of mass and luminosity in galaxies undergoing 
major mergers as a function of redshift.  We
find that the peak mass accretion rate per co-moving volume is 
$\sim 6 \times 10^{6}$ \solm Gpc$^{-3}$ year$^{-1}$ at $z \sim 1$. We also find
that the mass accretion rate per galaxy, $\dot{\rm M}_{\rm G}$ increases
as a function of redshift for all galaxies, as does the
fraction of mass in galaxies undergoing mergers.  For galaxies with 
M$_{\star} > 10^{9}$ \solm we find a maximum mass per galaxy accretion rate 
of $\dot{\rm M}_{\rm G} \sim 5.5 \times
10^{8}$ \solm Gyr$^{-1}$ at $z \sim 2.5$, with a rate evolution
given by $\dot{\rm M}_{\rm G} = 1.6 \times 10^{8}$ \solm Galaxy$^{-1}$ 
Gyr$^{-1}$ $(1+z)^{0.99\pm0.32}$.  The fraction of galaxy stellar mass density
involved in mergers also increases as a function of redshift, but much more
rapidly, and with a higher maximum fraction, for the brightest and most massive
systems.  For galaxies with M$_{\rm B} < -21$ or M$_{\star} > 10^{10}$ 
\solm, the fraction of mass involved in mergers is $\sim 0.5$ at
$z \sim 2.5$, demonstrating
that at least half of mass in the most massive galaxies in 
the nearby universe were involved in major mergers $\sim 9$ Gyrs ago.

\noindent v) Qualitative and quantitative comparisons of merger fractions
with results from Cold Dark Matter (CDM) simulations are in relatively good
agreement for the most massive systems at $z \sim 2.5$.  There is some 
discrepancy between the models and the observed
merger fractions for galaxies at low-masses and faint magnitudes,
especially at $z \sim 1$.  Our results are in agreement with the core result 
of CDM structure-formation models in which massive galaxies form and evolve
by merging.

Changes in the merger history through time can also explain a 
host of galaxy phenomenon which we have not considered or discussed in this 
paper, including: variations in cosmic star formation 
history (e.g. Madau et al. 1998), the peak in density of active
galactic nuclei (Boyle \& Terlevich 1998), the formation of black
holes (Menou, Haiman \& Narayanan 2001), and the evolution of supernovae and 
gamma-ray bursts.  In general,
our results are in agreement with the idea that major mergers have
occurred in large numbers in the past, and that a large fraction of
the most massive galaxies in the universe have formed by the
merging of lower mass systems.  While our results are for only a small
area of the sky, future observations with the Advanced Camera for Surveys
on the Hubble Space Telescope, such as the GOODS fields, will allow us to put 
firmer constraints on the
merging history of galaxies, including determining how the merger rate varies
as a function of environment.

A number of individuals have taken part in the development and acquisition of
the data used in this paper.  First and foremost, we thank the HDF-N NICMOS
team for their contributions and tremendous work obtaining the NIR data used 
in this paper, including acquiring spectroscopic and photometric
redshifts for galaxies in the HDF. We also thank Kirk Borne, Harry
Ferguson and Jay Gallagher for advice and 
suggestions regarding this work and Richard Ellis, Kevin Bundy and 
Mike Santos for comments on various drafts of this paper.  Andrew
Benson kindly supplied his semi-analytic results used here, which we greatly
appreciate. This work was supported by a National Science Foundation Astronomy
and Astrophysics Fellowship, and by NASA
HST Archival Researcher grant HST-AR-09533.04-A, both to CJC.
MAB acknowledges support from NASA grants GO-9126 from STScI, 
which is operated by AURA under contract NAS5-26555; from NASA/LTSA grant
NAG5-6032; and from NSF grant AST-9970780. MD and CP acknowledge support from 
NASA through grant number GO-07817.01-96A, also from the Space Telescope 
Science Institute.  Support from the University of 
Wisconsin-Madison, the Space Science Telescope Institute and the California 
Institute of Technology during the course of this work is also gratefully 
acknowledged.

\begin{figure}
\plotfiddle{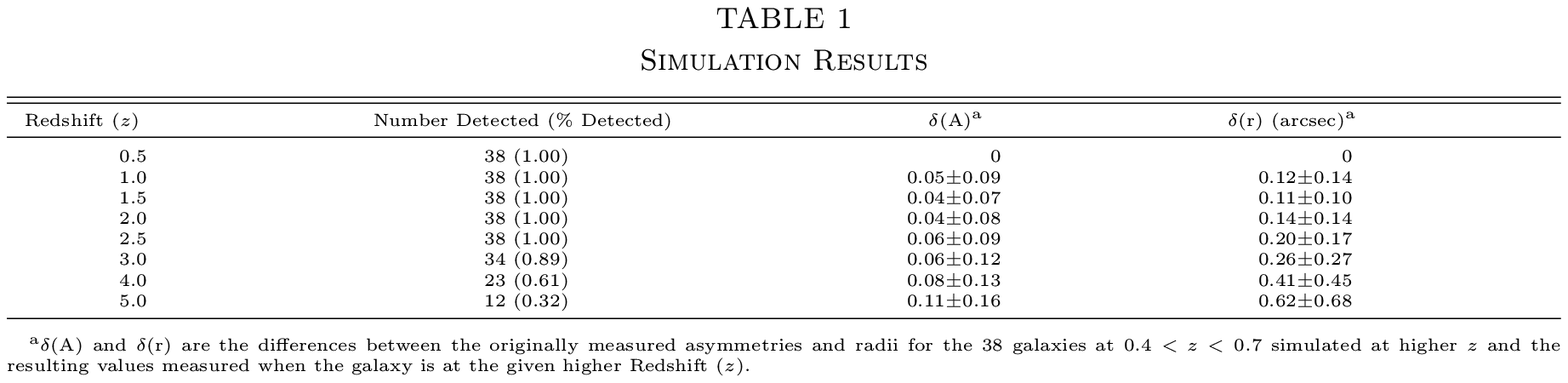}{6.0in}{0}{100}{100}{-310}{-170}
\end{figure}
 
\begin{figure}
\plotfiddle{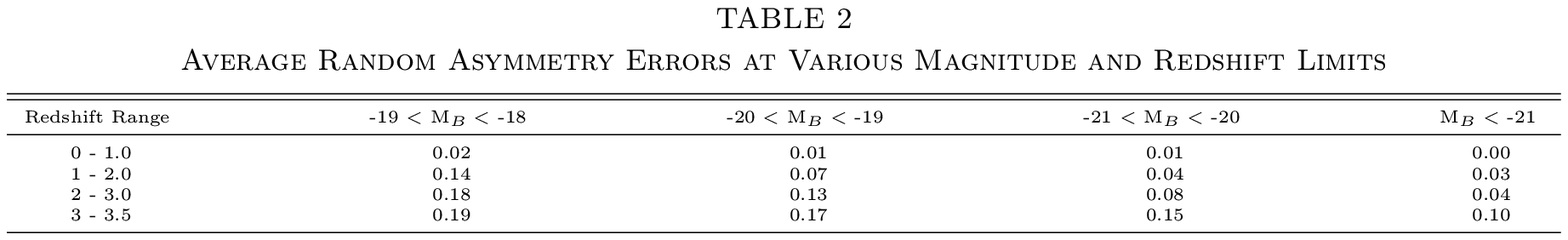}{6.0in}{0}{100}{100}{-310}{-170}
\end{figure}

\begin{figure}
\plotfiddle{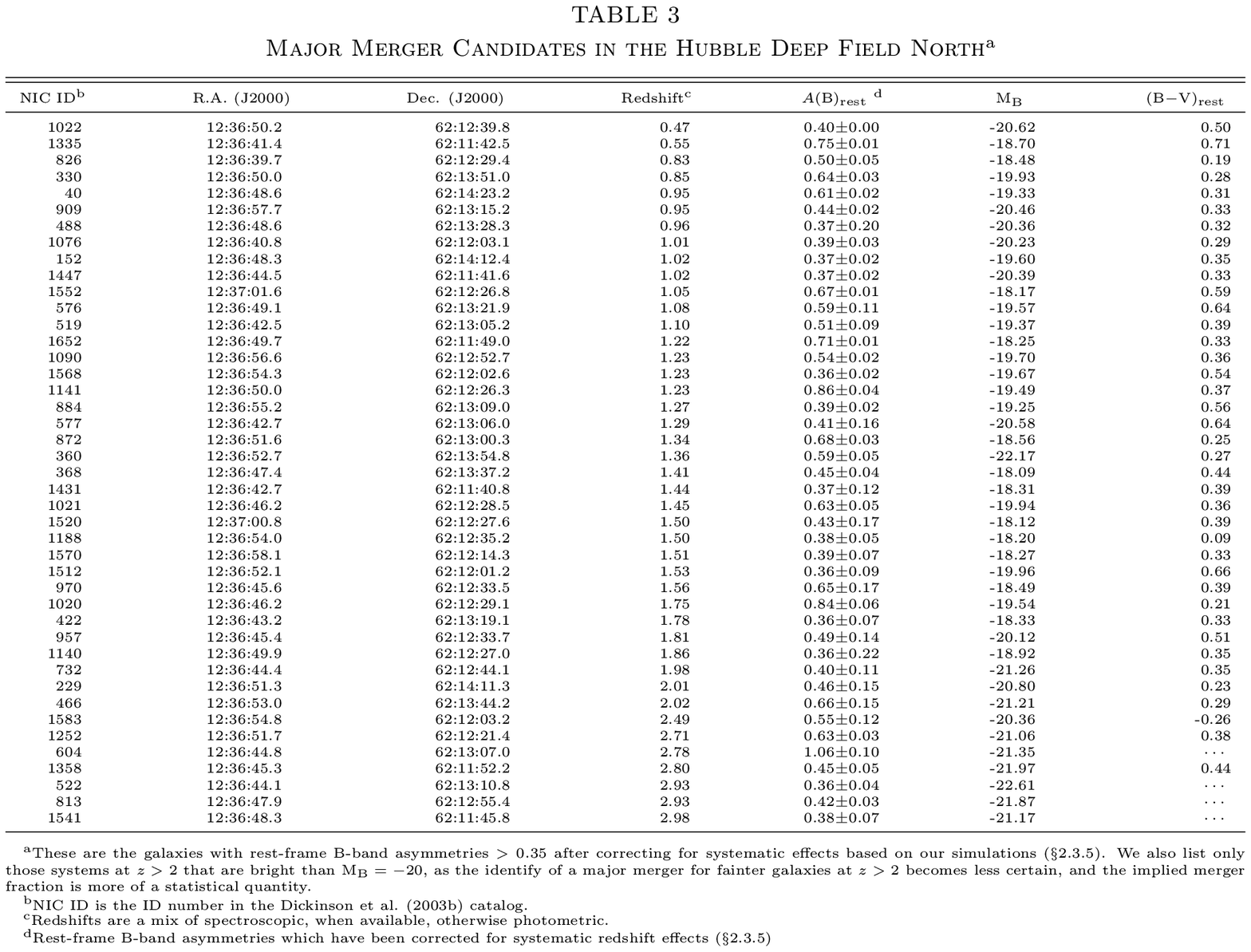}{6.0in}{0}{100}{100}{-310}{-170}
\end{figure}

\begin{figure}
\plotfiddle{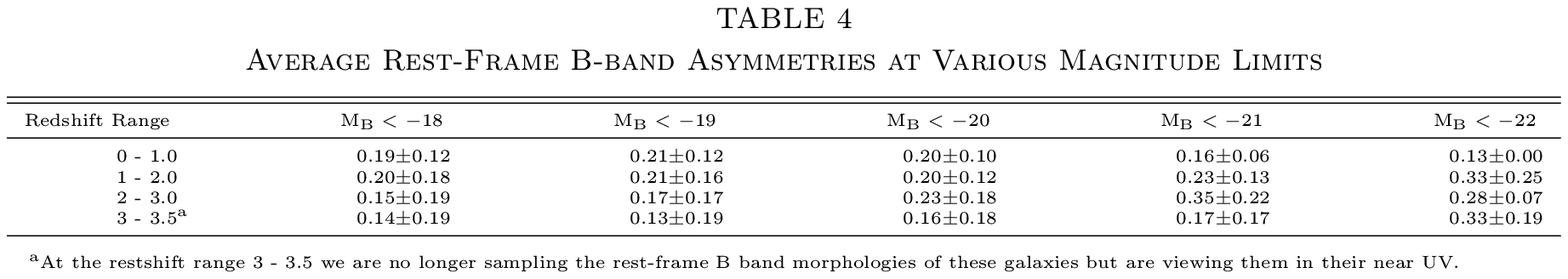}{6.0in}{0}{100}{100}{-310}{-170}
\end{figure}

\begin{figure}
\plotfiddle{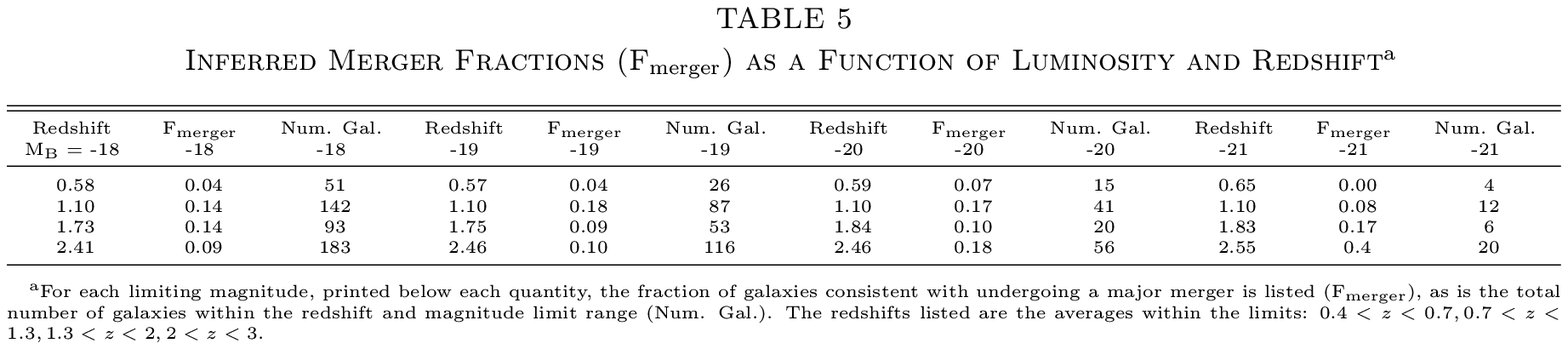}{6.0in}{0}{100}{100}{-310}{-170}
\end{figure}

\begin{figure}
\plotfiddle{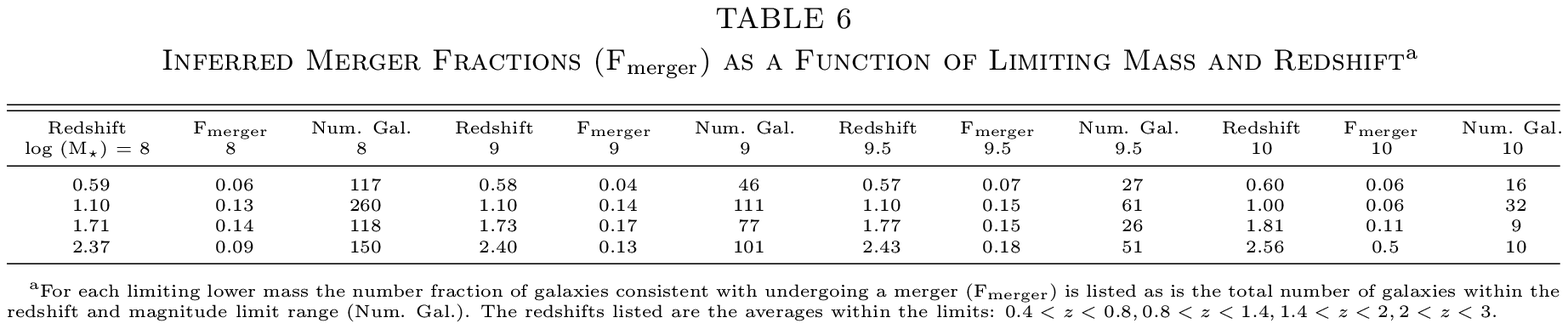}{6.0in}{0}{100}{100}{-310}{-170}
\end{figure}

\begin{figure}
\plotfiddle{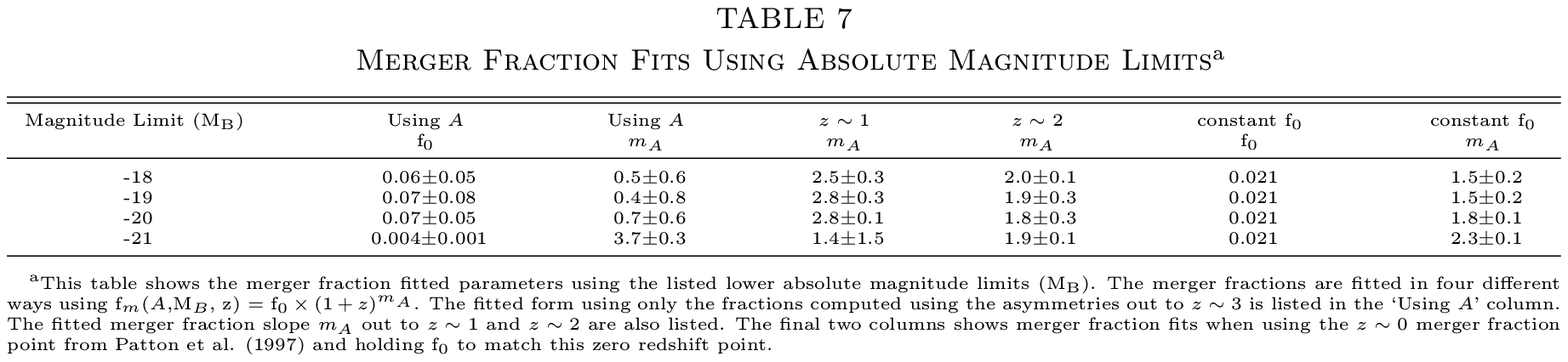}{6.0in}{0}{100}{100}{-310}{-170}
\end{figure}

\begin{figure}
\plotfiddle{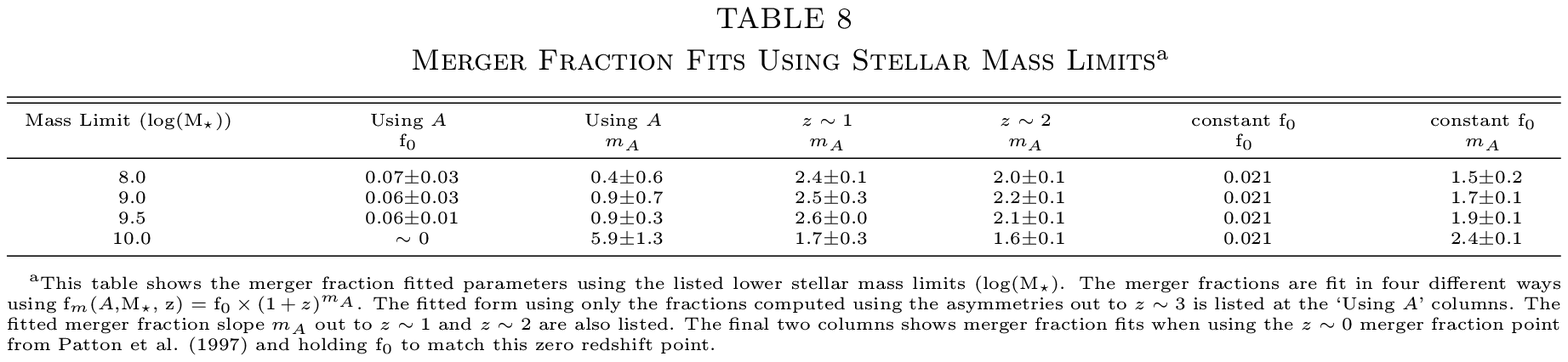}{6.0in}{0}{100}{100}{-310}{-170}
\end{figure}

\begin{figure}
\plotfiddle{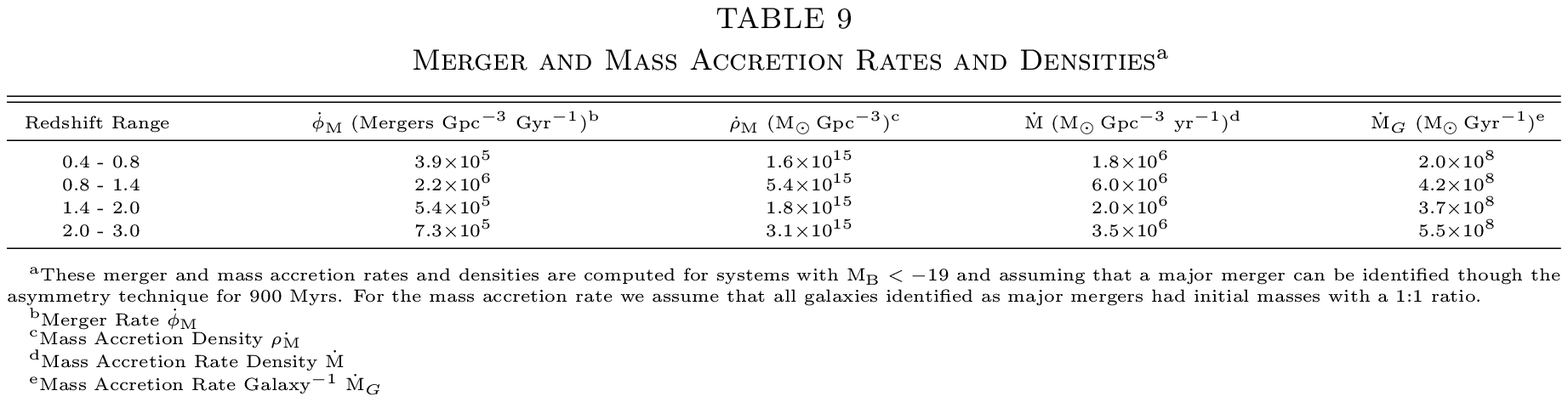}{6.0in}{0}{100}{100}{-310}{-170}
\end{figure}

\setcounter{figure}{0}

\begin{figure}
\plotfiddle{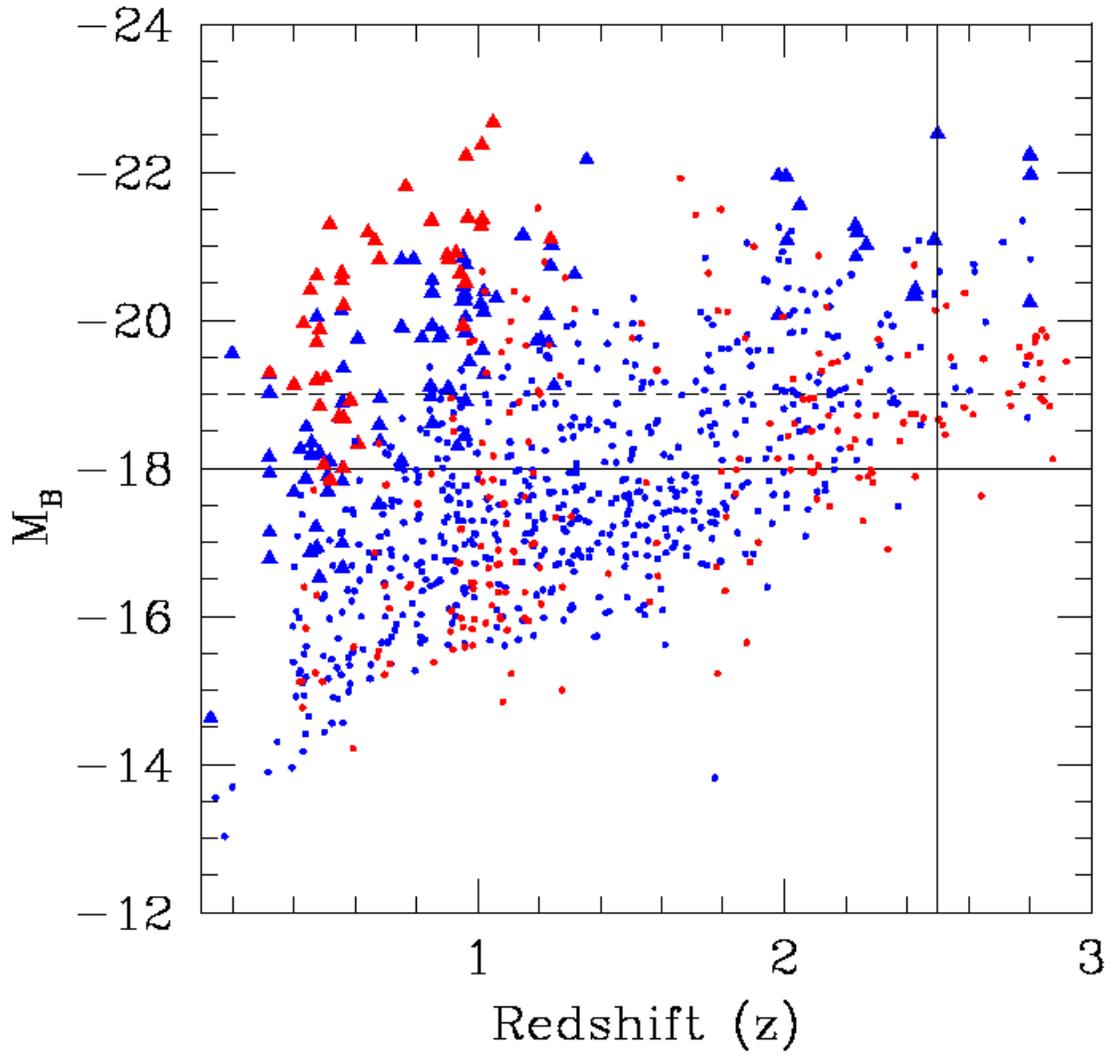}{6.0in}{0}{80}{80}{-250}{-100}
\caption{Plot of absolute magnitude vs. redshift for galaxies
in the Hubble Deep Field North.  The red symbols are for objects with
(B$-$V)$_{\rm rest} > 0.5$ and blue symbols are for (B$-$V)$_{\rm rest} < 0.5$.
The large symbols are objects with confirmed spectroscopic redshifts while
the smaller symbols are those objects with photometric redshifts.  The vertical
line at $z \sim 2.5$ is the limit we use for obtaining reliable morphological
parameters and galaxy detections, and the solid and dashed horizontal lines 
shows the M$_{\rm B} < -18$ and M$_{\rm B} < -19$ lower luminosity limits
we use throughout this paper (see text).}
\end{figure}
\clearpage

\begin{figure}
\plotfiddle{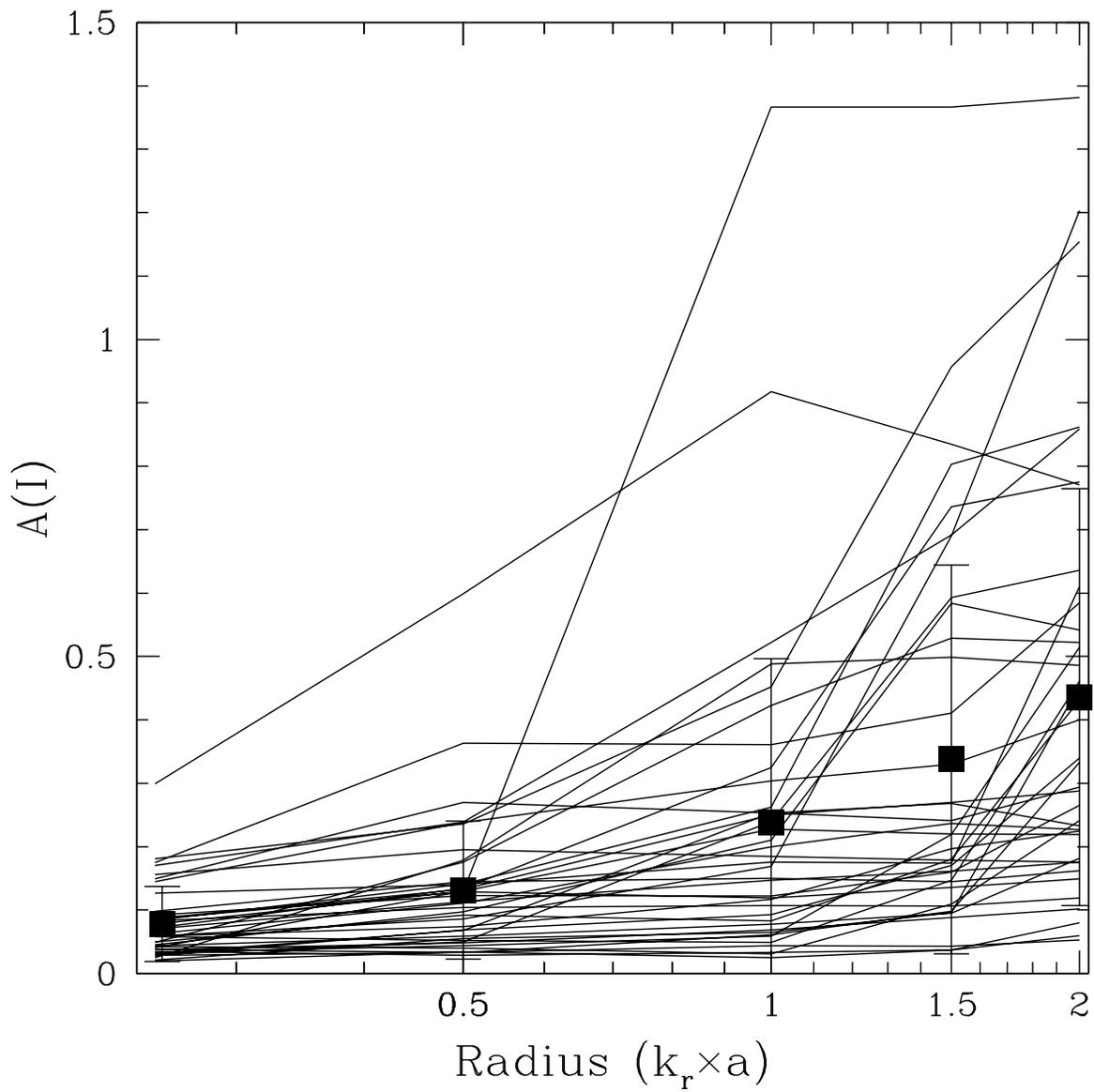}{6.0in}{0}{80}{80}{-250}{-100}
\caption{Plot of the asymmetries computed in the I band for
the 38 galaxies in the HDF with M$_{\rm B} < -18$ and at $0.4 < z < 0.7$ at
various fractions of the Kron multiplier (k$_{r}$) times semi-major axis 
($a$) radii. 
The solid box is the average asymmetry value at every radius, while errorbars
represent the 1 $\sigma$ variation of these averages.}
\end{figure}

\begin{figure}
\plotfiddle{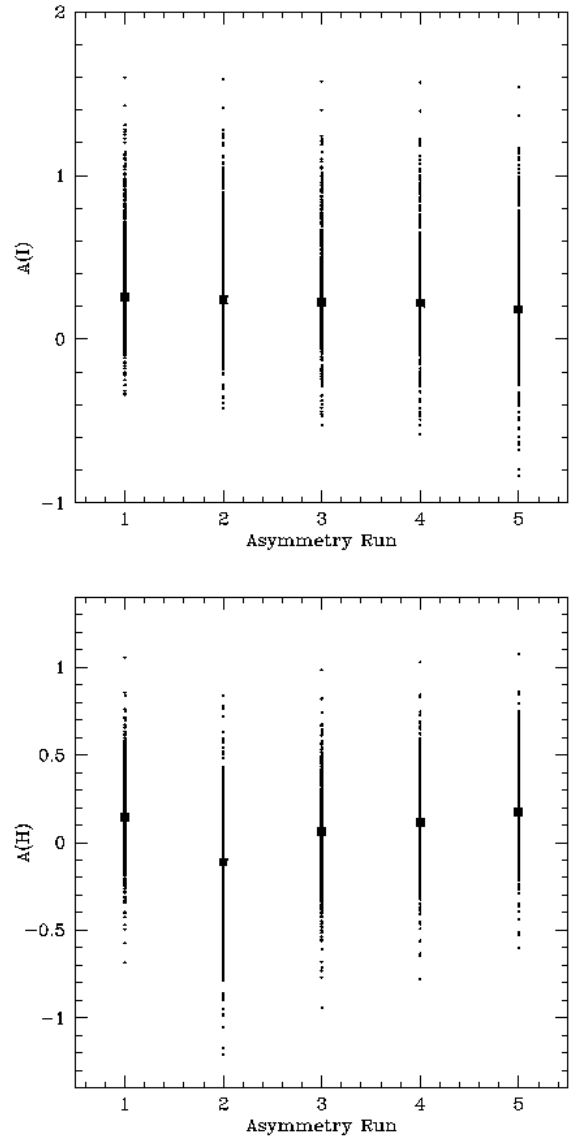}{6.0in}{0}{70}{70}{-200}{-100}
\vspace*{1in}
\caption{The distribution of asymmetries for all 1212 galaxies observed
in the HDF as measured in the F814W (I) band and the F160W (H) band after
using five different background positions for computing the background 
correction.}
\end{figure}
\clearpage

\begin{figure*}
\centerline{
\psfig{file=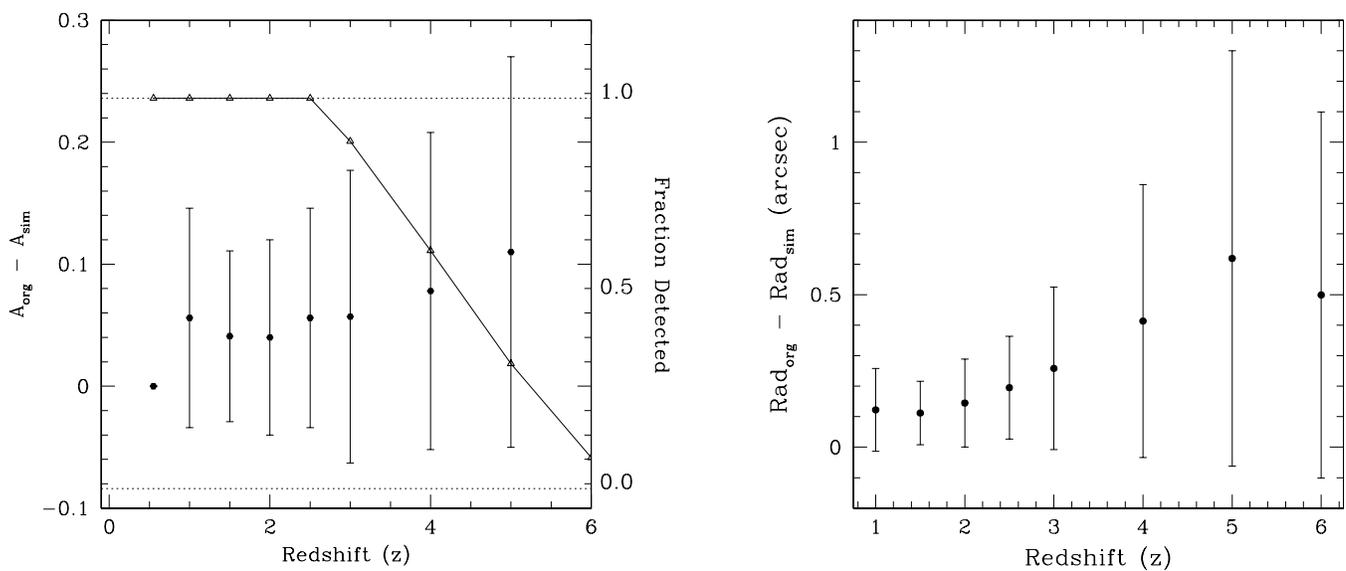,width=20cm,angle=-90}
}
\vspace*{-2cm}
\caption{Results of simulating the 38 HDF galaxies at 0.4 $< z < 0.7$ and
M$_{B} < -18$ to higher redshifts and then redoing the entire asymmetry
analysis from detection to measurement.  The left panel shows the average
and 1$\sigma$ variation of the asymmetry difference between the original
asymmetry and the simulated asymmetry ($A_{\rm org}$ - $A_{\rm sim}$).  Also
plotted on the right side of the left panel is the fraction of the 38
galaxies that are detected through SExtractor after performing the simulation.
The right panel shows the average and 1$\sigma$ variations of the difference,
in arcseconds, between the original radii measured (Rad$_{\rm org}$) and the
simulated radii (Rad$_{\rm sim}$).
}
\label{a68c4}
\end{figure*}


\begin{figure}
\plotfiddle{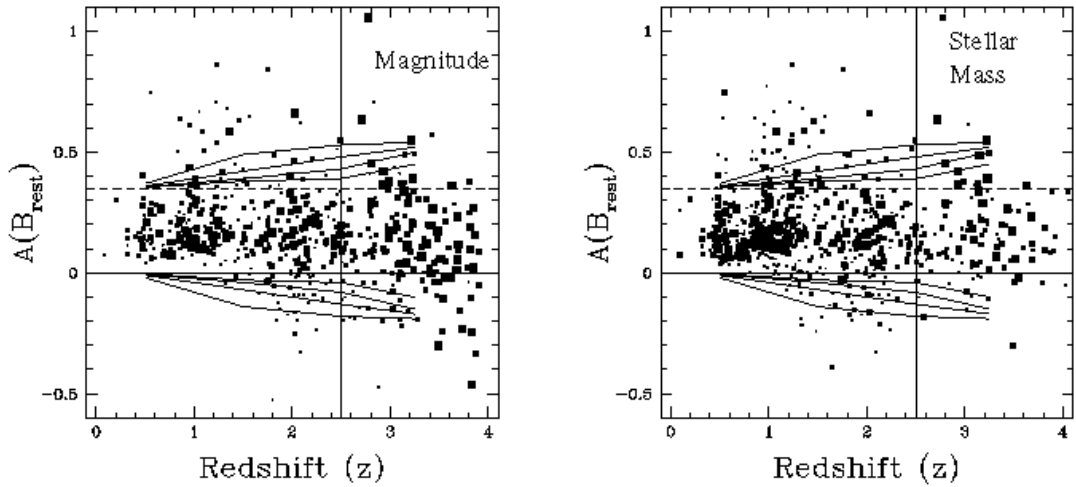}{6.0in}{0}{80}{80}{-250}{-100}
\vspace*{-1in}
\caption{Plot of the rest frame B-band 
asymmetry measurements $A$(B$_{\rm rest}$)
as a function of redshift ($z$).  The size of the plotted point is 
proportional to (left) the luminosity (magnitude) of the galaxy and (right) 
the stellar mass of the galaxy.  The smallest points are
objects with a. $-18 >$ M$_{\rm B} > -19$, b. $8 <$ M$_{*} < 9$.  The 
larger symbols are for galaxies with a. $-19 >$ M$_{\rm B} > -20$, 
b. $9 <$ M$_{*} < 9.5$, and a. $-20 >$ M$_{\rm B} > -21$, b. 
$9.5 <$ M$_{*} < 10$.  The largest symbols are for galaxies with
a. $-21 >$ M$_{\rm B} > -23$ and b. M$_{*} > 10$. The solid line is 
$A$(B$_{\rm rest}$) = 0
and the dashed line is the limit for mergers at $A$(B$_{\rm merger}$) = 0.35.  
The solid diagonal lines originating at  A(B$_{\rm rest}$) = 0 and 
A(B$_{\rm merger}$) = 0.35 show the average random error on the
asymmetry index as a function of redshift and limiting magnitude, such
that each line is either $A(z) = 0.35$ + error($z$) or $A(z) = 0-$error$(z)$.  
These lines are for galaxies at magnitudes, from the nearest to their 
respecitve horizontal line outward, M$_{\rm B} < -21$,
$-21$ $<$ M$_{\rm B} < -20$, $-20$ $<$ M$_{\rm B} < -19$, and
$-19 <$ M$_{\rm B} < -18$.}
\end{figure}
\clearpage

\begin{figure*}
\centerline{
\psfig{file=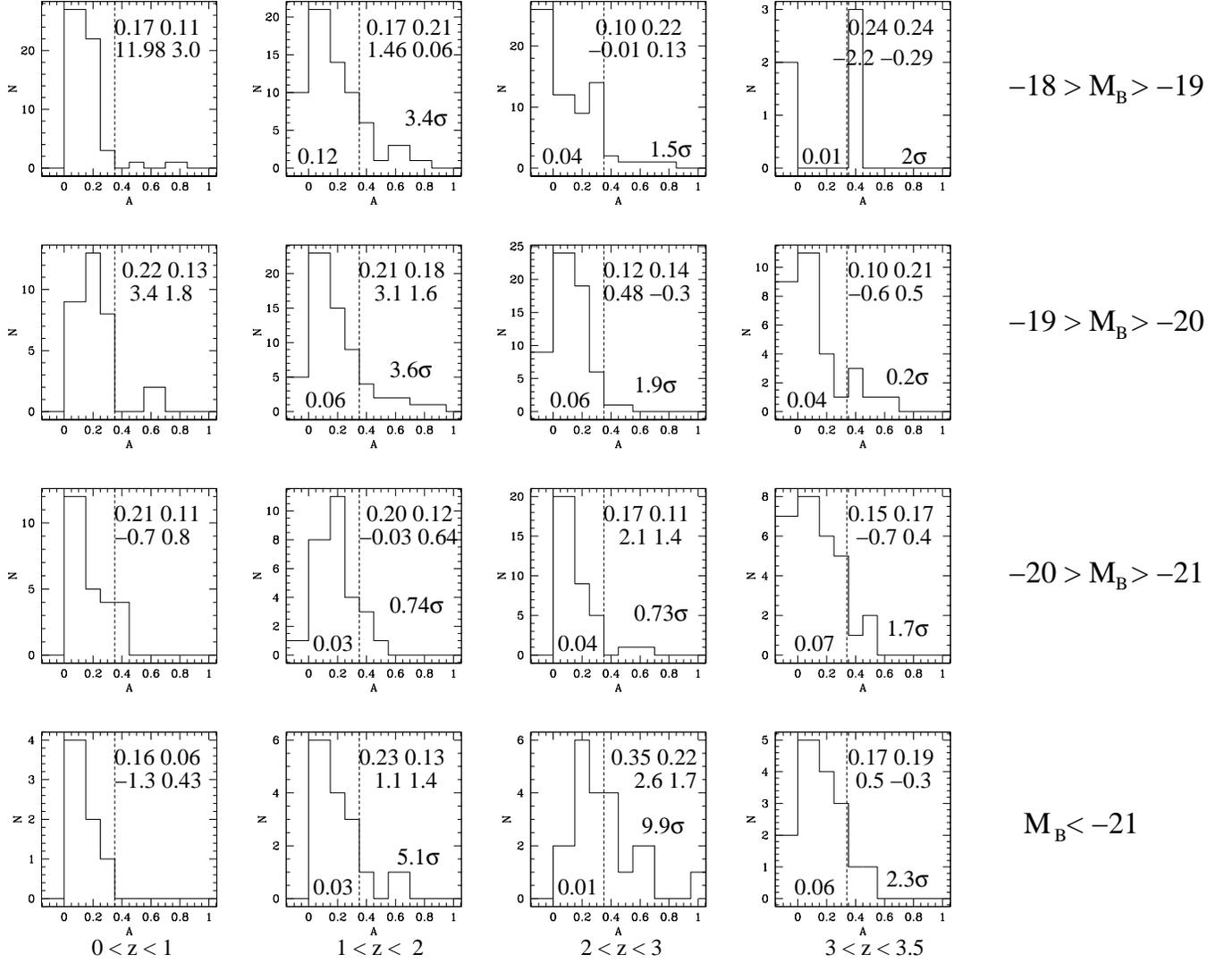,width=20cm,angle=-90}
}
\vspace*{0cm}
\caption{Distribution of asymmetries as a function of absolute
magnitude and redshift.  The top four numbers are, clockwise top-left, 
the mean 
asymmetry, its 1 $\sigma$ variation, its skewness and kurtosis values.  
The bottom left
number is the difference in the average random error for the galaxies in
that particular magnitude and redshift bin and the previous lower redshift
bin at the same magnitude.  The bottom right number is the significance, 
based on Monte Carlo simulations discussed in \S 3.2, that an increase
in asymmetries are not due to an increase in random errors.
The dashed vertical line shows the $A_{\rm merger}$ limit.}
\label{a68c4}
\end{figure*}


\clearpage

\begin{figure}
\plotfiddle{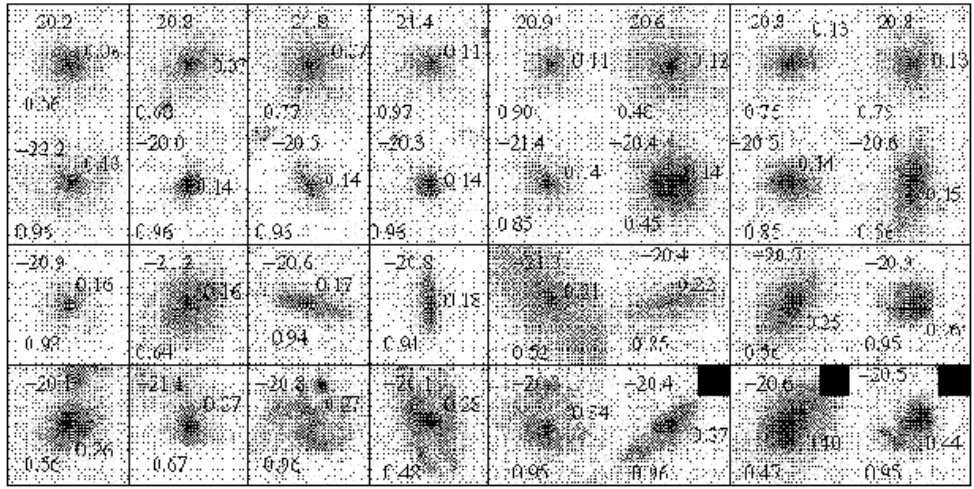}{6.0in}{0}{70}{70}{-200}{-100}
\vspace*{-1in}
\caption{Images of all galaxies in the Hubble Deep Field North with 
M$_{\rm B} < -20$, between 0 $< z <$ 1, ordered by increasing asymmetry, as
seen in F814W.  The upper number in each panel is
the M$_{\rm B}$ of each galaxy and the bottom number is its redshift, while
the number on the right hand side is the galaxy's $A(\rm B)$ value. Galaxies
consistent with being mergers based on their asymmetries have a solid
dark box in the right hand corner of their respective panels.}
\end{figure}

\setcounter{figure}{6}

\clearpage
\begin{figure}
\plotfiddle{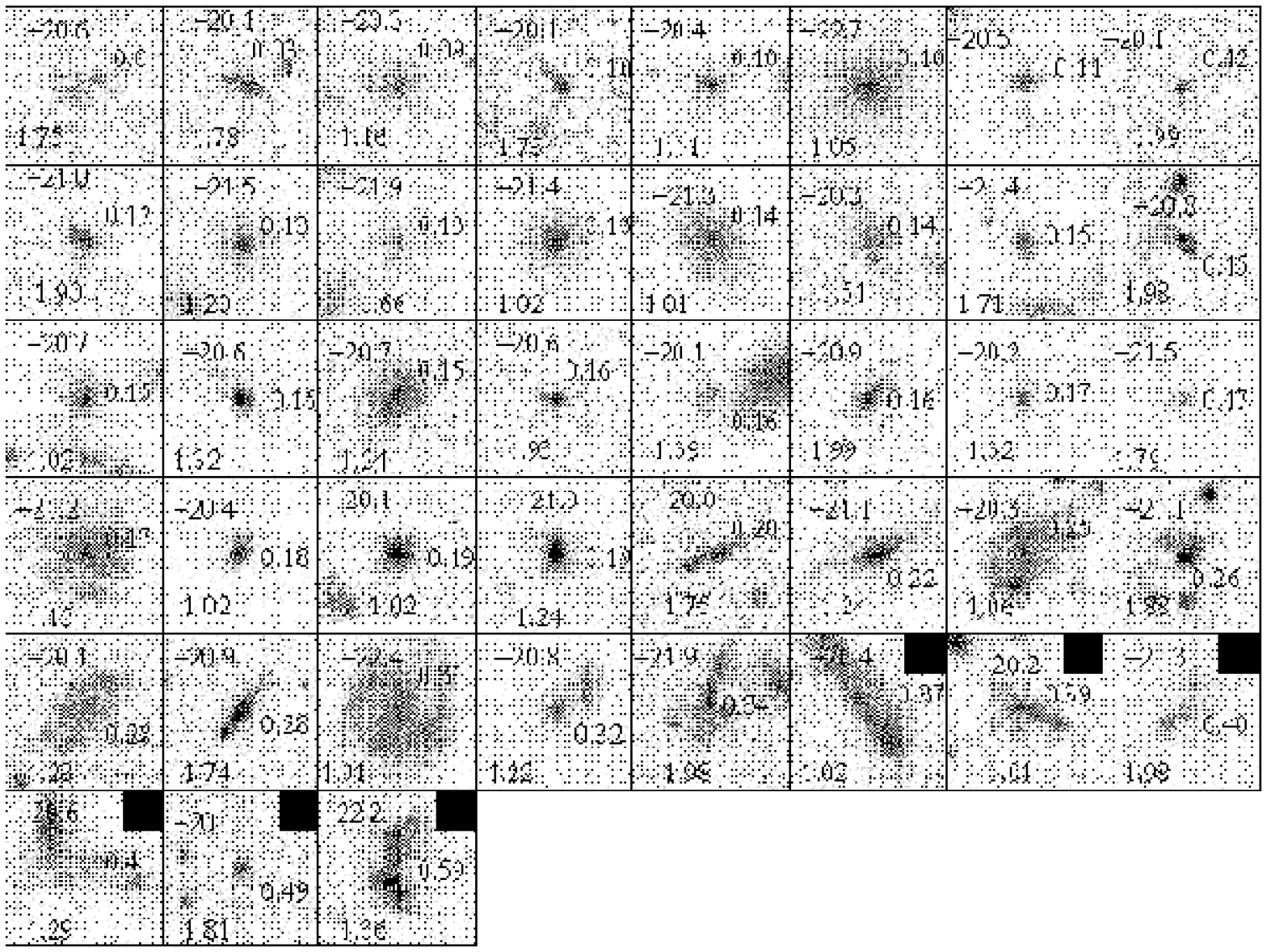}{6.0in}{0}{70}{70}{-200}{-100}
\caption{continued - same as part a. expect all galaxies with
M$_{\rm B} < -20$, between 1 $< z <$ 2 are shown.}
\end{figure}

\setcounter{figure}{6}

\clearpage
\begin{figure}
\plotfiddle{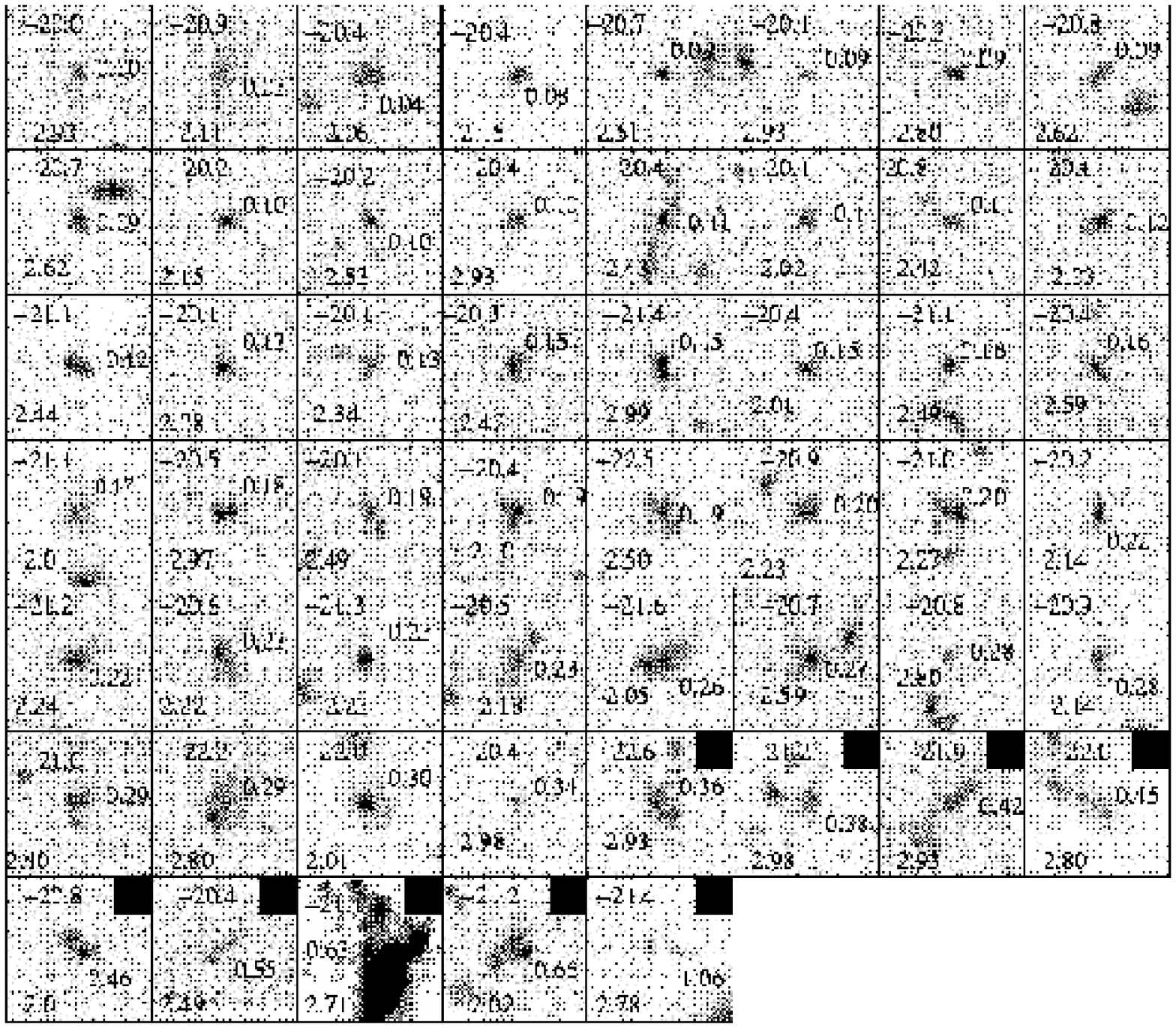}{6.0in}{0}{70}{70}{-200}{-100}
\caption{continued - same as part a. expect all galaxies with
M$_{\rm B} < -20$, between 2 $< z <$ 3 are shown.}
\end{figure}

\begin{figure}
\plotfiddle{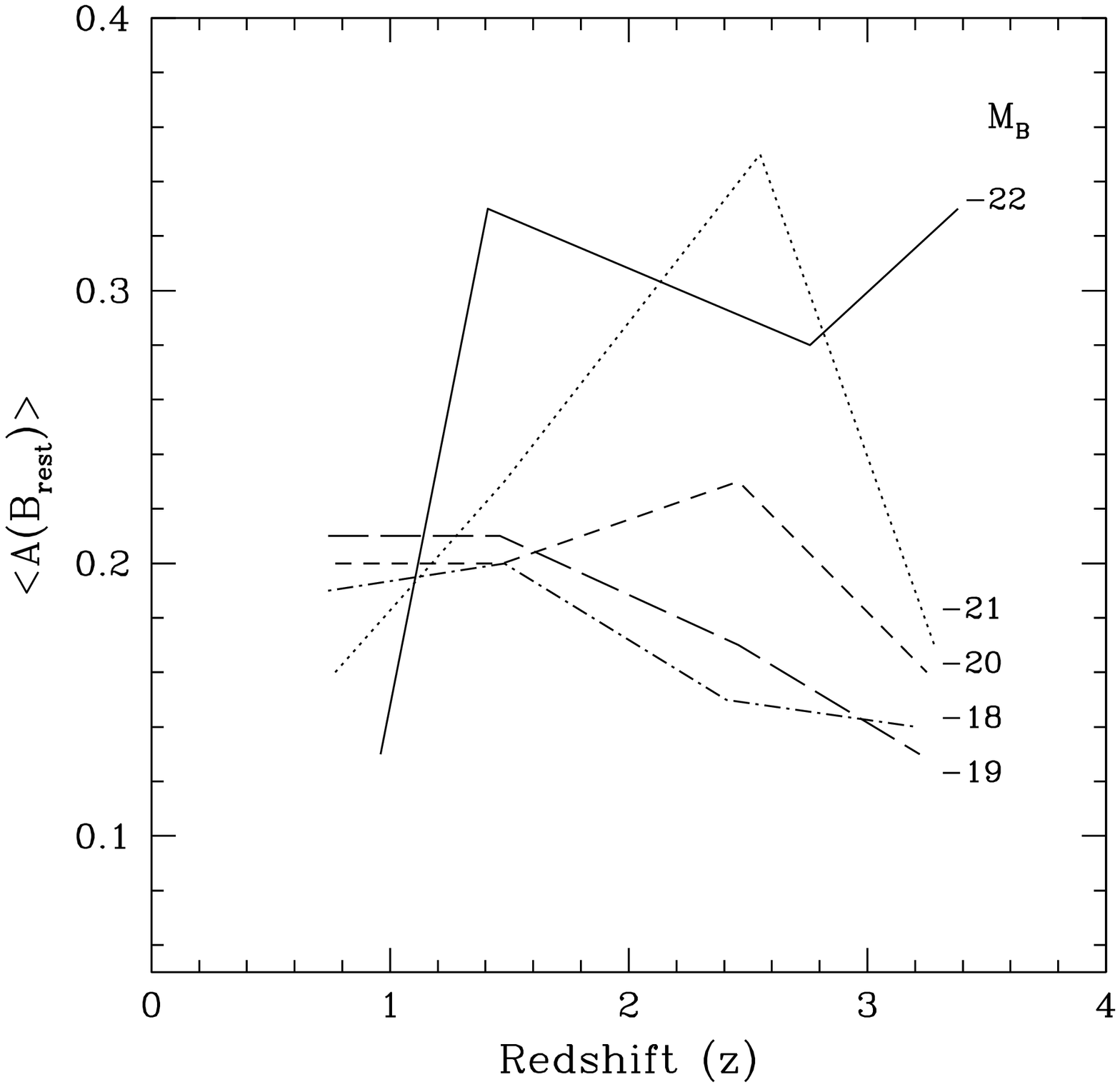}{6.0in}{0}{80}{80}{-250}{-100}
\caption{Plot of the average rest frame B-band asymmetry for
galaxies at different magnitudes limits as a function of redshift.  Variations
of these averages are listed in Table~4.}
\end{figure}
\clearpage

\begin{figure}
\plotfiddle{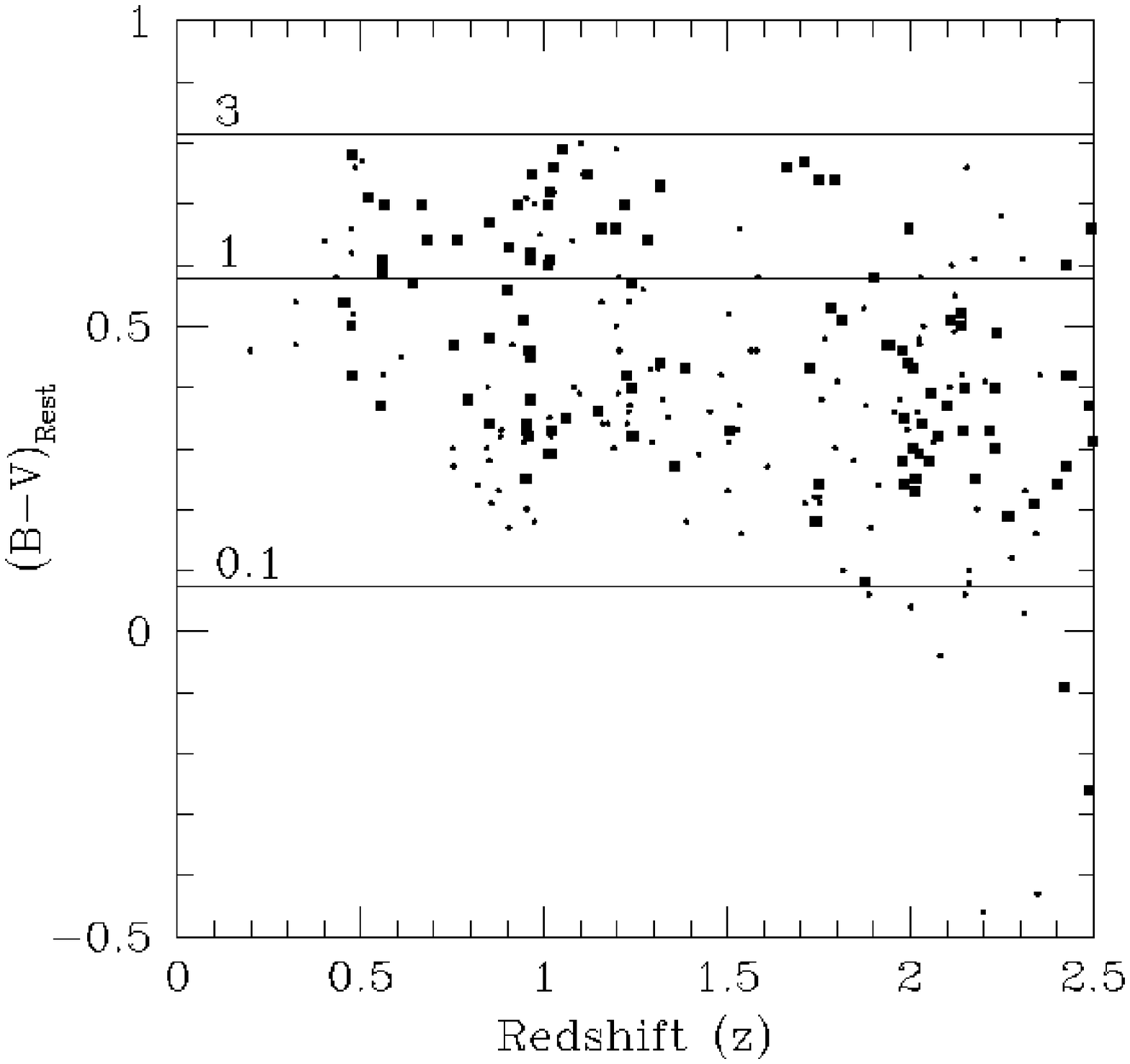}{6.0in}{0}{80}{80}{-250}{-100}
\caption{The (B$-$V) color distribution of galaxies plotted as a function of
redshift ($z$).  The larger boxes are galaxies at M$_{\rm B} < -20$ while
the small dots are galaxies with M$_{\rm B} > -20$. The solid lines
show synthesis stellar population colors with ages of 0.1, 1 and 3 Gyr 
for systems that formed in bursts.}
\end{figure}
\clearpage

\begin{figure*}
\centerline{
\psfig{file=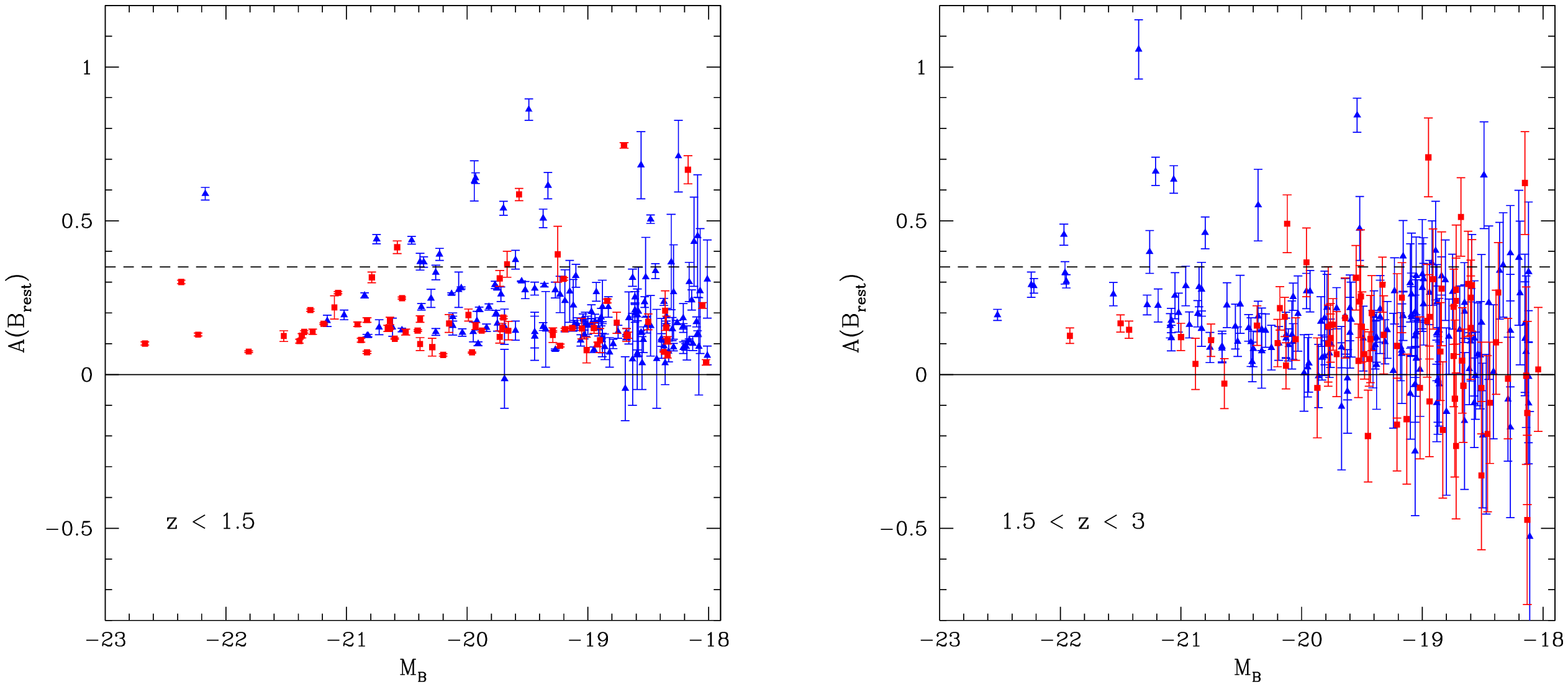,width=20cm,angle=-90}
}
\vspace*{-2cm}
\caption{Relationship between the rest frame B-band asymmetry 
$A$(B$_{\rm rest}$)
and absolute magnitude (M$_{\rm B}$) for HDF galaxies at two different
redshift bins: $z < 1.5$ and $1.5 < z < 3$.  The color of each point
is dictated by the rest frame (B$-$V) color of each galaxy, such that
blue points are systems with (B$-$V) $<$ 0.5 and red points are for
galaxies with (B$-$V) $> 0.5$.}
\label{a68c4}
\end{figure*}


\begin{figure}
\plotfiddle{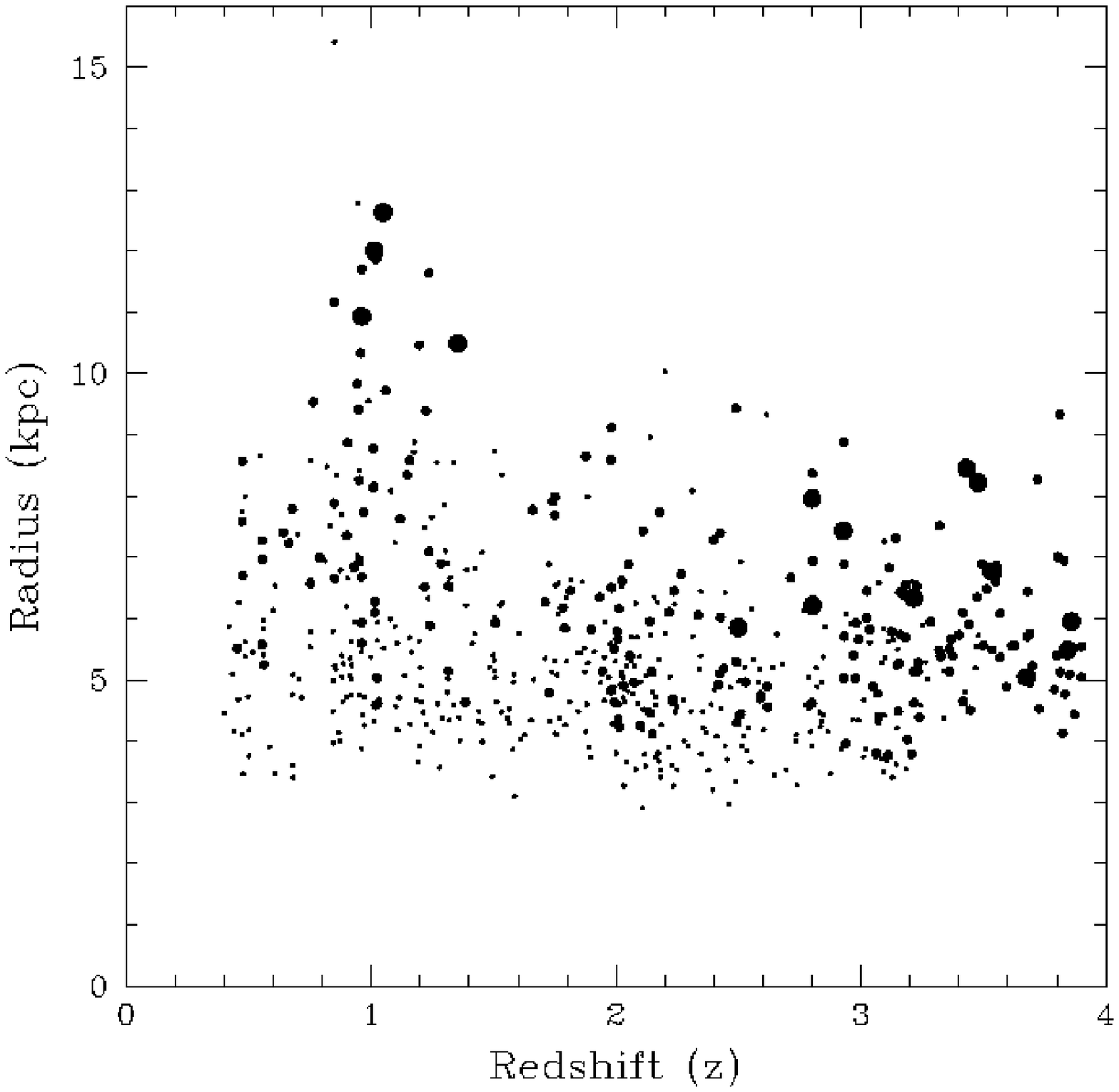}{6.0in}{0}{80}{80}{-250}{-100}
\caption{The distribution of HDF galaxy radii ($0.5 \times {\rm k}_{r} \times$
a), in the rest-frame B-band to $z < 2.5$ and observed H-band at
$z > 2.5$, corrected for
redshift effects (\S 2.3.5 \& Table~1) plotted as a function of redshift 
($z$).   The size of the symbol is proportional to luminosity, such that the 
smallest symbols are for galaxies with $-20 <$ M$_{\rm B} < -18$ and the two
larger symbols are for systems with $-22 <$ M$_{\rm B} < -20$ and
M$_{\rm B} < -22$.}
\end{figure}
\clearpage

\begin{figure}
\plotfiddle{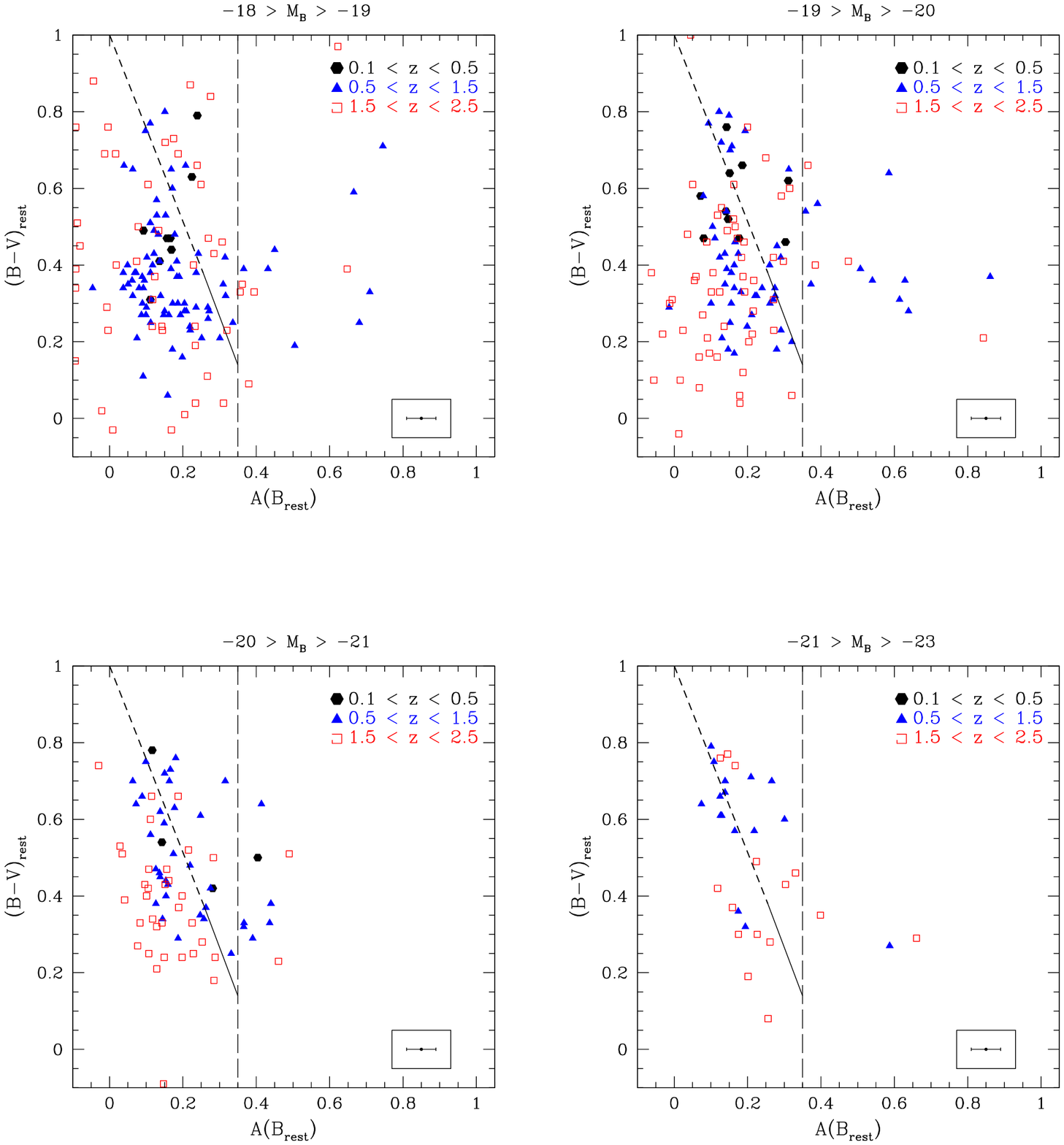}{6.0in}{0}{70}{70}{-200}{-100}
\vspace*{1in}
\caption{The rest-frame (B$-$V) color verses asymmetry diagram plotted at
four different luminosity intervals listed at the top of each panel.  
Each panel also plots galaxies 
divided into different redshift ranges at each luminosity. 
The diagonal dashed line is the relationship between (B-V) color
and the rest-frame B-band asymmetry $A$(B$_{\rm rest}$) found for
nearby normal galaxies (Conselice et al. 2000a).  This line is extrapolated
to bluer colors by the solid diagonal line.  The vertical long-dashed
line shows the $A_{\rm merger}$ $=$ $A$(B$-$V = 0.4) + 3$\sigma ($B$-$V) limit
we use for identifying mergers (see text). }
\end{figure}

\clearpage

\clearpage
\begin{figure}
\plotfiddle{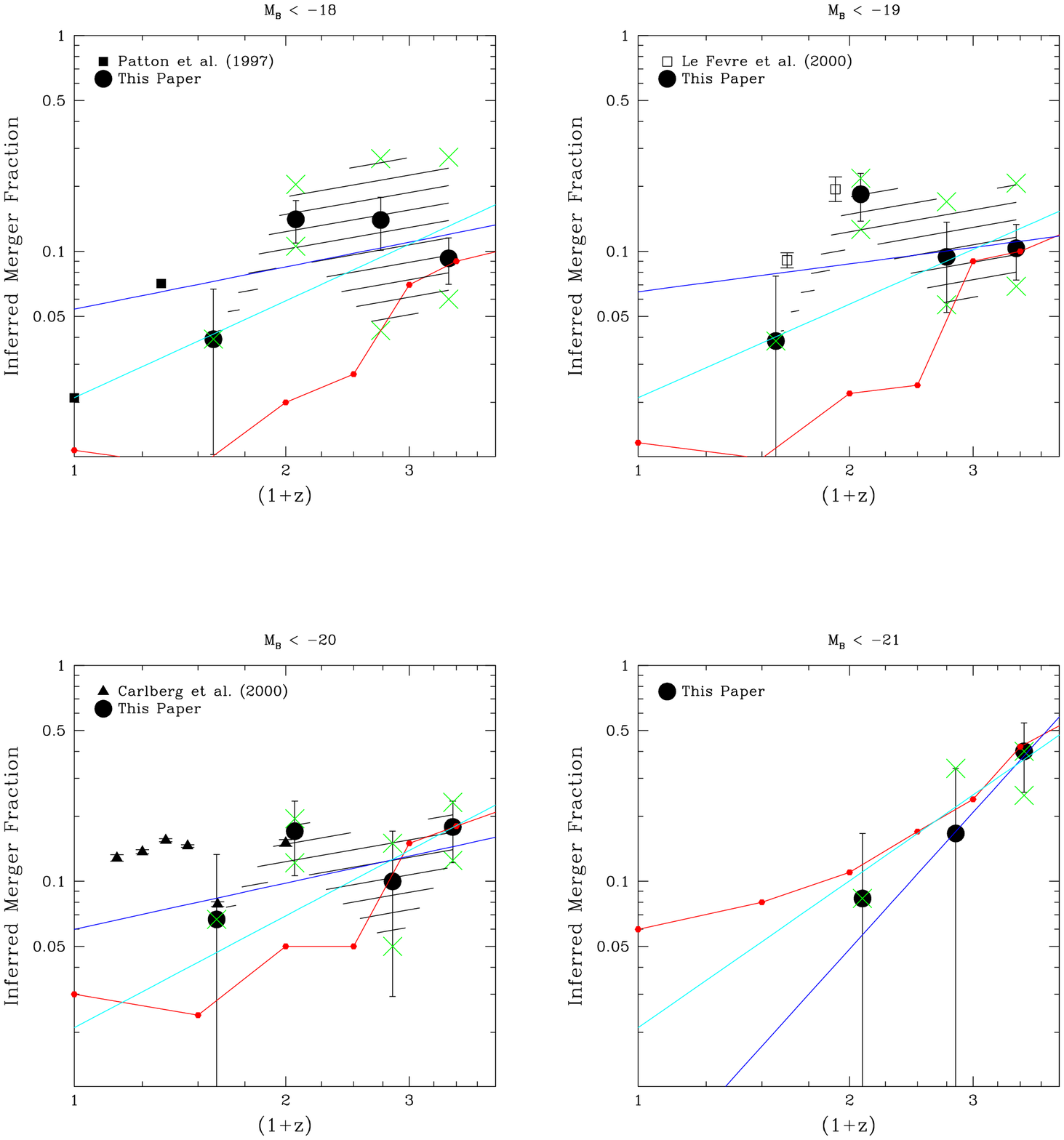}{6.0in}{0}{70}{70}{-200}{-100}
\vspace*{1in}
\caption{Plots of the evolution of merger fractions as a function of
redshift.  The solid large round symbols are merger fractions computed
using the asymmetry technique.  The green crosses are merger fractions
computed after using the +$1\sigma$ and -$1\sigma$ values for each
galaxy's asymmetry.  The other symbols are defined in the legend printed
on each panel, and are merger fractions
found from galaxy pairs by Patton et al. (1997), LeFevre et al. (2000)
and Carlberg et al. (2000).  The straight blue line shows the merger
fraction fit in the form f = f$_{0} \times (1+z)^{m_{A}}$ when only
using merger fractions computed by using the asymmetry derived fractions, and
the cyan line is this fit when using the asymmetry merger fractions
and holding f$_{0}$ = 0.021, the $z \sim 0$ value found by Patton et al. 
(1997). The solid red line and points are merger fractions selected
in an analogous way from the semi-analytic CDM simulation results of Benson
et al. (2002).}
\end{figure}
\clearpage

\begin{figure}
\plotfiddle{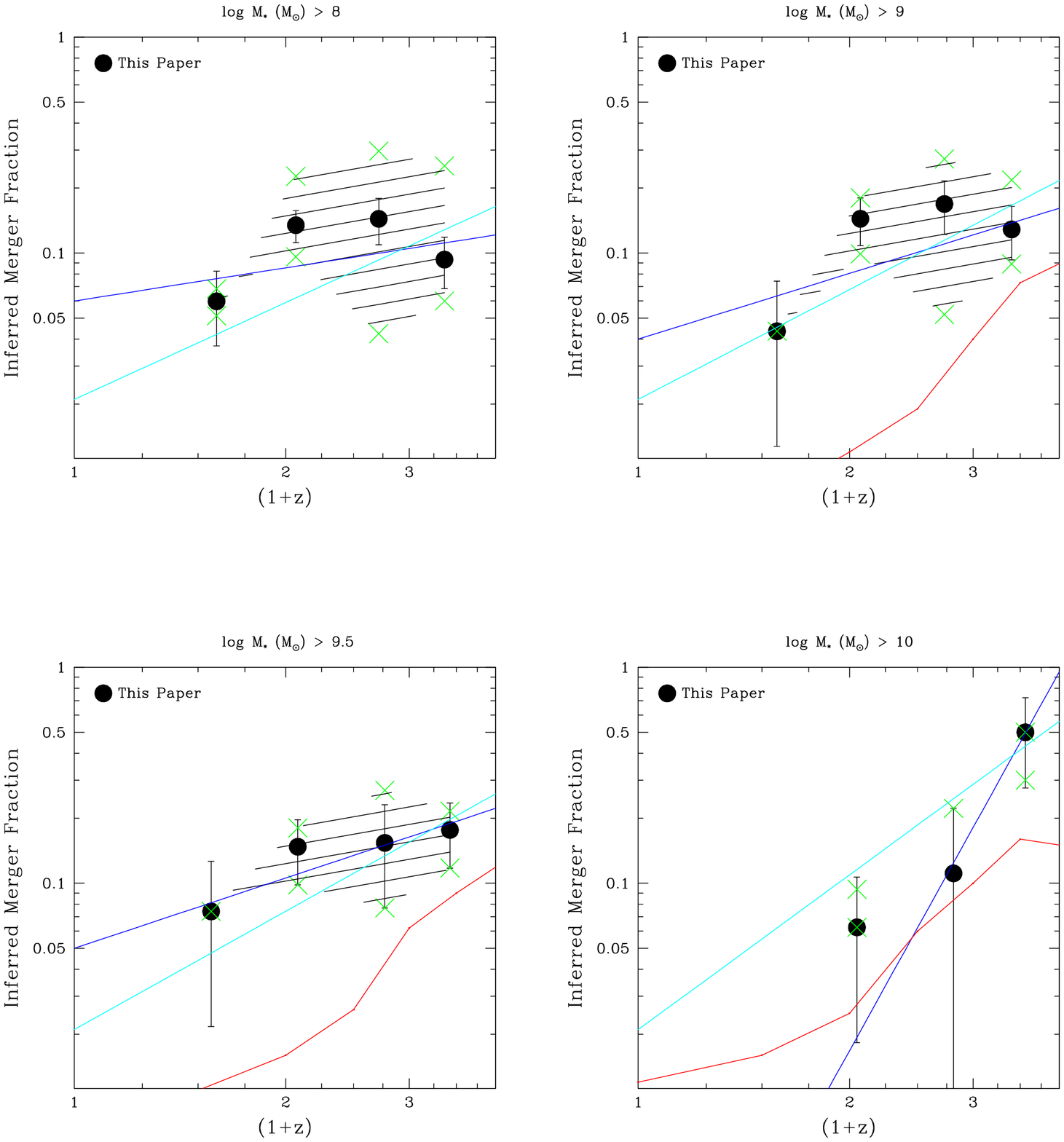}{6.0in}{0}{70}{70}{-200}{-100}
\vspace*{1in}
\caption{An analogous plot of Figure~13 except that the merger fractions
are selected based on stellar mass limits, rather than absolute
magnitudes.  }
\end{figure}
\clearpage

\begin{figure}
\plotfiddle{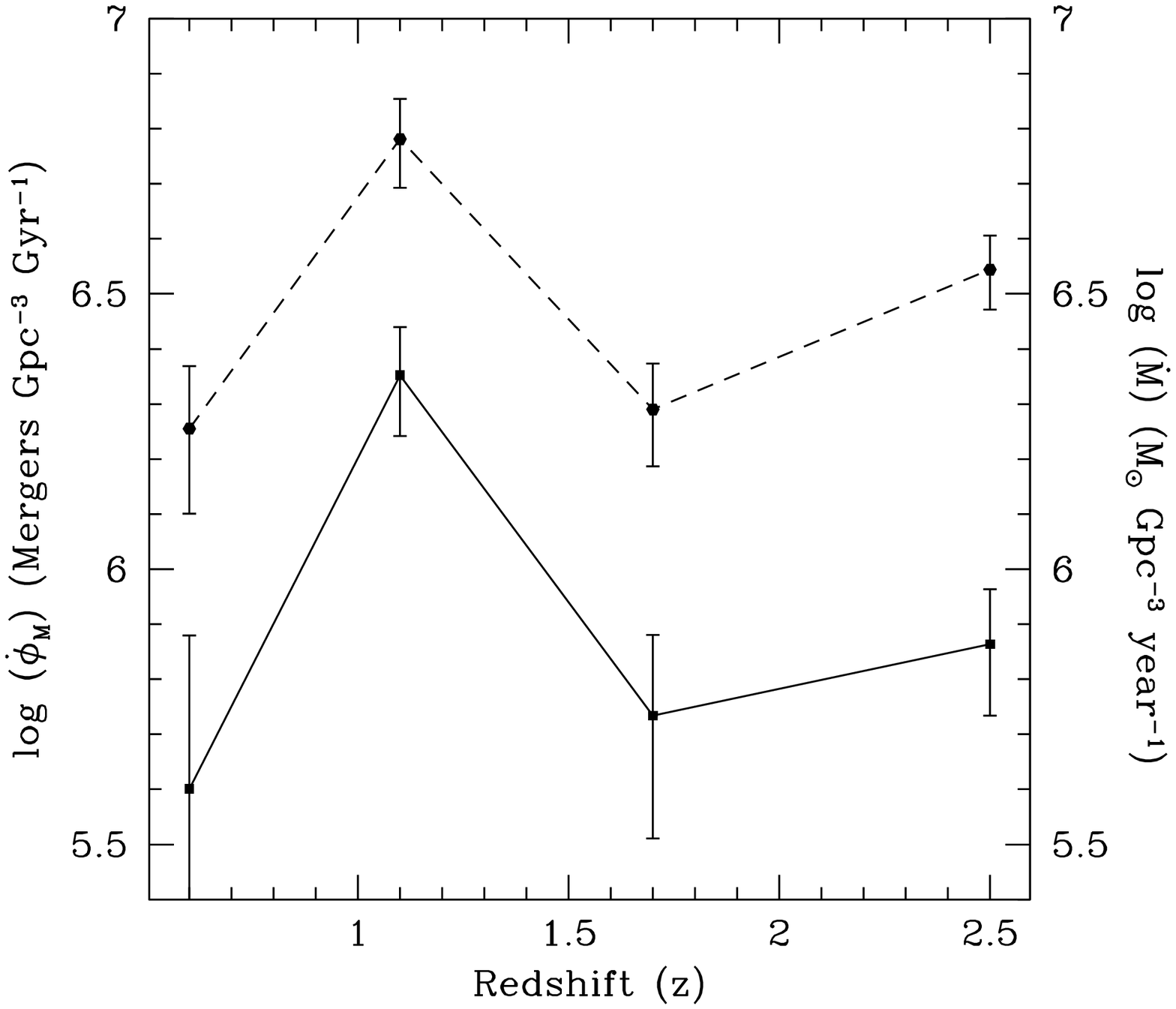}{6.0in}{0}{80}{80}{-250}{-100}
\caption{The merger rate $\dot{\phi}_{M}$ in units of number of mergers 
Gpc$^{-3}$ Gyr$^{-1}$ plotted as a function of redshift ($z$) (solid line)
and the stellar mass accretion rate $\dot{\rm M}$ in units of \solm Gpc$^{-3}$ 
year$^{-1}$ (dashed line) for galaxies at M$_{\rm B} < -19$.}
\end{figure}
\clearpage

\begin{figure}
\plotfiddle{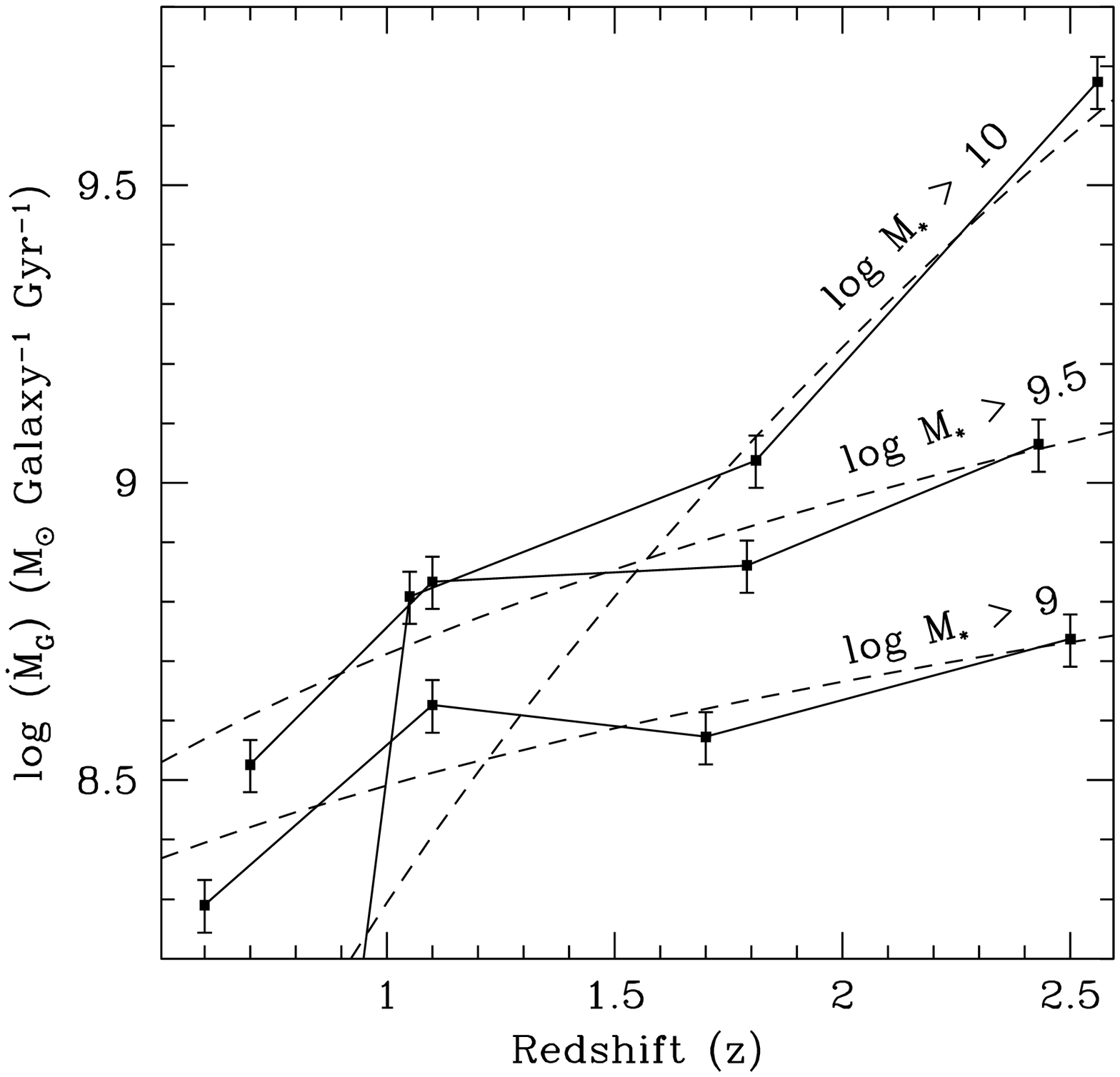}{6.0in}{0}{80}{80}{-250}{-100}
\caption{The stellar mass accretion rate per galaxy $\dot{\rm M}_{\rm G}$, in
units of \solm~Galaxy$^{-1}$ Gyr$^{-1}$ plotted as
a function of redshift ($z$) for galaxies with stellar masses
greater than 10$^{9}$, 10$^{9.5}$ and 10$^{10}$ \solm.  The dashed
lines are fits to the accretion rates listed in equations 1-3.}
\end{figure}
\clearpage

\begin{figure}
\plotfiddle{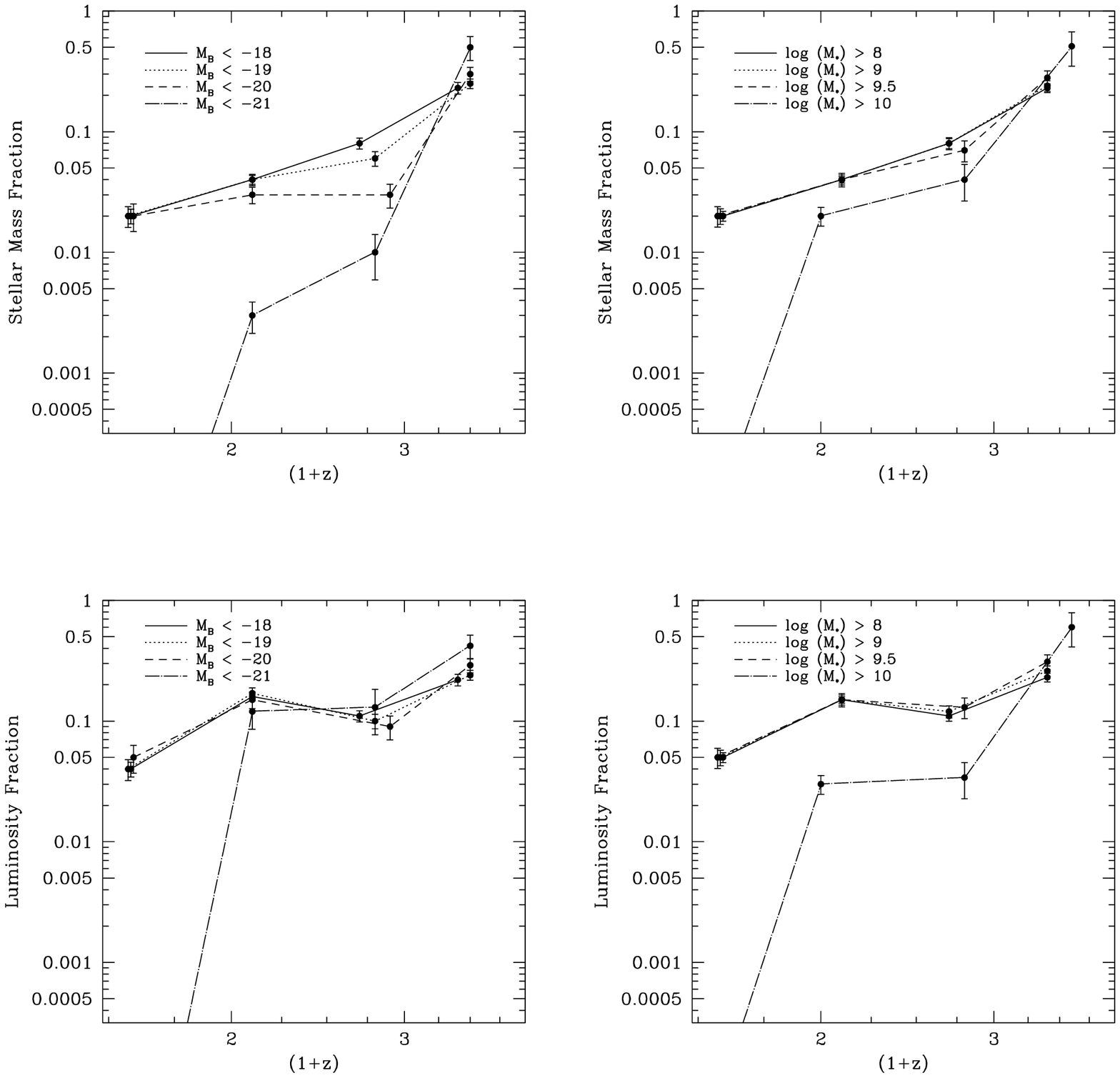}{6.0in}{0}{80}{80}{-250}{-100}
\vspace*{0.5in}
\caption{Plots of the fraction of stellar mass and galaxy luminosity involved 
in mergers as a function of redshift and magnitude (left panels) and stellar 
mass (right panels). The different lines show the fraction of
mass and rest-frame B-band luminosity involved in mergers at different
luminosity and stellar mass upper limits. }
\end{figure}
\clearpage

\end{document}